\documentclass[twocolumn,showpacs,showkeys,amsmath,amssymb,pra]{revtex4}    

\usepackage{bm}
\usepackage{color}
\usepackage{graphicx}   
\newcommand{\rd}{{\mathrm d}}
\newcommand{\re}{{\mathrm e}}
\newcommand{\ri}{{\mathrm i}}
\newcommand{\rt}{(\bm r;t)} 
\newcommand{\ph}{\widehat{\psi}}
\newcommand{\phd}{\widehat{\psi}^\dagger}
\newcommand{\pH}{\widehat{\psi}_{\rm H}^{\phantom\dagger}}
\newcommand{\pHd}{\widehat{\psi}_{\rm H}^{\dagger}}
\newcommand{\Hh}{\widehat{H}}
\newcommand{\Uh}{\widehat{U}}
\newcommand{\vac}{| {\rm vac} \rangle}
\newcommand{\wP}{\widetilde{\Psi}}
\newcommand{\aL}{a_1^{\phantom{\dagger}}}
\newcommand{\aR}{a_2^{\phantom{\dagger}}}
\newcommand{\aj}{a_j^{\phantom{\dagger}}}
\newcommand{\aLd}{a_1^\dagger}
\newcommand{\aRd}{a_2^\dagger}
\newcommand{\ajd}{a_j^\dagger}
  
\begin{document}

\title[Gross-Pitaevskii equation]
      {$N$-coherence vs.\ $t$-coherence: \\ 
      An alternative route to the Gross-Pitaevskii equation}

\author{Bettina Gertjerenken}
\author{Martin Holthaus}

\affiliation{Institut f\"ur Physik, Carl von Ossietzky Universit\"at, 
	D-26111 Oldenburg, Germany}

                  
\date{August 04, 2015}

\begin{abstract}
We show how a candidate mean-field amplitude can be constructed from the exact 
wave function of an externally forced $N$-Boson system. The construction 
makes use of subsidiary $(N-1)$-particle states which are propagated in time 
in addition to the true $N$-particle state, but does not involve spontaneous 
breaking of the $U(1)$ symmetry associated with particle number conservation. 
Provided the flow in Fock space possesses a property which we call maximum
stiffness, or $t$-coherence, the candidate amplitude actually satisfies the 
time-dependent Gross-Pitaevskii equation, and then serves as macroscopic wave 
function of the forced $N$-particle system. The general procedure is 
illustrated in detail by numerical calculations performed for the model of 
a driven bosonic Josephson junction, which allows one to keep track of all 
contributions which usually are subject to uncontrolled assumptions. These 
calculations indicate that macroscopic wave functions can persist even under 
conditions of strong forcing, but are rapidly destroyed upon entering a regime 
of chaotic dynamics. Our results provide a foundation for future attempts to 
manipulate, and actively control, macroscopic wave functions by means of 
purposefully designed force protocols.      
\end{abstract} 

\pacs{03.75.Kk, 03.75.Lm, 67.85.De}


\keywords{Macroscopic wave function, mean-field approximation, coherence,
		bosonic Josephson junction, nonequilibrium quantum dynamics,
		strongly driven quantum systems, quantum chaos}

\maketitle 


\section{Introduction and Overview}
\label{sec:1}

This paper concerns the time evolution of a system of $N$ interacting identical
spinless Bose particles of mass~$m$, where $N$ is assumed to be large. At each 
moment~$t$ in time it is described by a Schr\"odinger wave function $\Psi$ 
which is symmetric in all its spatial arguments,
\begin{equation}
	\Psi(\bm r_1, \ldots, \bm r_N; t) = 
	\Psi(\bm r_{P(1)}, \ldots, \bm r_{P(N)}; t)
\label{eq:MBW}
\end{equation}
for each permutation~$P$ of the indices $1,\ldots,N$. Its dynamics are
governed by an $N$-particle Schr\"odinger equation
\begin{eqnarray}
	& & 
	\ri \hbar \frac{\partial}{\partial t} 
	\Psi(\bm r_1, \ldots, \bm r_N; t) 
\\	& = &
	\left[ \sum_{j=1}^N h_1(\bm r_j;t) 
	+ \frac{1}{2} \sum_{j \neq k} U(\bm r_j, \bm r_k) \right]
	\Psi(\bm r_1, \ldots, \bm r_N; t) \; , 
\nonumber
\end{eqnarray}	
where  
\begin{equation}
	h_1\rt = -\frac{\hbar^2}{2m}\Delta + V\rt 
\label{eq:SPH}
\end{equation}
denotes a single-particle Hamiltonian containing an external potential~$V$ 
which may depend explicitly on time, so as to exert some controlling influence
on the system, and the interparticle interaction~$U$ naturally is symmetric,
\begin{equation}
	U(\bm r, \bm r') = U(\bm r', \bm r) \; .	
\label{eq:TPI}
\end{equation}
We will assume that $U$ can be replaced by an effective contact interaction 
with strength~$g$,
\begin{equation}
	U_{\rm eff}(\bm r, \bm r') = g \, \delta(\bm r - \bm r') \; ,
\label{eq:COP}
\end{equation}				
which means that we restrict ourselves to the low-energy 
regime~\cite{HuangYang57}. 

The central question addressed in this work is under what conditions an 
approximate description of the dynamics in terms of a mean-field amplitude 
$\Phi$ evolving according to the time-dependent Gross-Pitaevskii 
equation~\cite{Gross61,Pitaevskii61,Gross63,Gardiner97,CastinDum98,Leggett01,
PethickSmith08,PitaevskiiStringari03} is viable, and, if so, how that amplitude
is related to the actual wave function~$\Psi$. Since $\Phi = \Phi\rt$ depends 
on only one spatial coordinate $\bm r$, it contains much less information than 
the many-body wave function~(\ref{eq:MBW}) could possibly carry. Therefore, 
a reduction of the full $N$-particle dynamics to the mean-field level without 
essential loss of information is feasible only if already the $N$-particle 
wave function itself is relatively {\em simple\/}, or {\em ordered\/}, so that 
the mean-field amplitude, if it exists, may be regarded as an order parameter 
of the system~\cite{Leggett00}. 
According to common knowledge deriving from London's theory of 
superfluids~\cite{London64} the mean-field amplitude should correspond to
a ``macroscopic wave function'', that is, to a macroscopically occupied
single-particle orbital. This is fully in line with the preservation of
information: In the ideal case where all $N$~particles occupy the same 
orbital, no information  is lost if $\Phi\rt$ equals that orbital. But this
leads to further questions when the system is subjected to time-dependent
forcing: Even if the initial state is given exactly by an $N$-fold occupied
single-particle orbital, this initial order might be destroyed by the external
force, and the time-dependent many-particle state may become arbitrarily
complicated, no longer admitting a mean-field description. On the other hand,
one may formally solve the Gross-Pitaevskii equation with an arbitrary type of 
external forcing incorporated into the single-particle Hamiltonian~$h_1$.
Hence, there must be some sort of indicator which tells one whether or not 
the solution to the time-dependent Gross-Pitaevskii equation can actually 
serve as a macroscopic wave function of the forced $N$-particle system. Here 
we explore this intuitive idea in mathematical terms.    

As a guideline for the following deliberations, and to state the subject as 
clearly as possible,  we briefly sketch a popular pseudo-derivation of the 
Gross-Pitaevskii equation. To this end, we introduce the usual Fock space 
annihilation and creation operators $\ph(\bm r)$ and $\phd(\bm r)$ which obey 
the canonical Bose commutation relations 
\begin{eqnarray}
	\left[ \ph(\bm r) , \ph(\bm r') \right] & = & 0
\nonumber \\
	\left[ \ph^\dagger(\bm r) , \ph^\dagger(\bm r') \right] & = & 0		
\nonumber \\
	\left[ \ph(\bm r) , \ph^\dagger(\bm r') \right] & = & 
	\delta(\bm r - \bm r') \; ,
\label{eq:BCR}
\end{eqnarray}
so that the Fock-space many-body Hamiltonian takes the form~\cite{Fock32}
\begin{eqnarray}
\label{eq:MBH}
	\Hh(t) & = & \int \! \rd^3 r \, 
	\ph^\dagger(\bm r) h_1\rt \ph(\bm r)
\\ 	& + & 
	\frac{1}{2} \int \! \rd^3 r \! \int \! \rd^3 r' \, 
	\ph^\dagger(\bm r) \ph^\dagger(\bm r') U(\bm r, \bm r')
	\ph(\bm r') \ph(\bm r) \; .
\nonumber
\end{eqnarray}
Here we use the ``hat''-symbol to designate operators acting in Fock space, 
in contrast to single-particle operators such as the one given by 
Eq.~(\ref{eq:SPH}).

With the help of the unitary operator $\Uh(t,t_0)$ which effectuates the
time evolution from some initial moment~$t_0$ to $t$ and thus obeys the
equation 
\begin{equation}
	\ri \hbar \frac{\rd}{\rd t} \Uh(t,t_0) = \Hh(t) \Uh(t,t_0) \; , 
\end{equation}
the field operator $\ph(\bm r)$ is transformed to the Heisenberg picture
through the familiar prescription  
\begin{equation}
	\pH\rt = \Uh^\dagger(t,t_0) \ph(\bm r) \Uh(t,t_0) \; , 
\end{equation}
leading to an equation of motion of the form 
\begin{equation}
	\ri \hbar \frac{\rd}{\rd t} \pH\rt = 
	\Uh^\dagger(t,t_0) \left[ \ph\rt, \Hh(t) \right] \Uh(t,t_0) \; .
\end{equation}
Working out the commutator appearing on the right-hand side, one obtains
\begin{eqnarray}
\label{eq:WOC}
	\ri \hbar \frac{\rd}{\rd t} \pH\rt 
	& = & h_1\rt \pH\rt
\\	
	& + & \!\!\!\! \int \! \rd^3 r' \, \pHd(\bm r';t) U(\bm r, \bm r')
	\pH(\bm r';t) \pH\rt \; ;
\nonumber
\end{eqnarray}
inserting the contact potential~(\ref{eq:COP}), one is left with
\begin{eqnarray}
\label{eq:HOM}
	\ri \hbar \frac{\rd}{\rd t} \ph_{\rm H}\rt
	& = & h_1\rt \ph_{\rm H}\rt
\\	
	& + & g \, \ph^\dagger_{\rm H}\rt \ph_{\rm H}\rt \ph_{\rm H}\rt \; .
\nonumber
\end{eqnarray}
Now comes a major assumption: The mean-field amplitude $\Phi$ is supposed
to be given by the expectation value of the Heisenberg field operator, 
\begin{equation}
	\sqrt{N} \Phi\rt := 
	\left\langle \ph_{\rm H}\rt \right\rangle \; .	
\label{eq:SBS}	
\end{equation}
This definition requires clarification. If it is taken literally, each 
$N$-particle state $| \Psi \rangle_N$, no matter how ``simple'', can only 
produce a left-hand side equal to zero, since the  $(N-1)$-particle state 
$\ph_{\rm H}\rt | \Psi \rangle_N$ is orthogonal to $| \Psi \rangle_N$ in 
Fock space. Hence, when working with Eq.~(\ref{eq:SBS}) one supposes that 
the superselection rule for the total particle number~$N$ is somehow 
violated, and the state $|\Psi\rangle$ with respect to which the expectation 
values~(\ref{eq:SBS}) are taken is some coherent superposition of $N$-particle 
states $|\Psi\rangle_N$ with different~$N$,
\begin{equation}
	| \Psi \rangle = \sum_N a_N | \Psi \rangle_N \; .
\label{eq:SUP}
\end{equation}
This corresponds to the idea that the $U(1)$ symmetry associated with particle 
number conservation is spontaneously broken.
While that concept may be helpful for simplifying formal calculations, 
it remains an auxiliary device not rooted on the grounds of the actually 
given $N$-particle system. This deficiency has led to several profound 
discussions in the literature, and to the development of number-conserving 
approaches~\cite{Gardiner97,CastinDum98,Leggett01,GirardeauArnowitt59,
Girardeau98,GardinerMorgan07}. The unreflected use of the notion of
spontaneous symmetry breaking has been severely challenged by Leggett, 
who criticizes that {\em there are no circumstances in which Eq.~(\ref{eq:SUP}) 
is the physically correct description of the system\/}, and maintains that the 
definition~(\ref{eq:SBS}) is ``liable to generate pseudoproblems, and is best 
avoided''~\cite{Leggett01}. Thus, one of the goals of the present study is to 
provide an alternative, logically more transparent definition of the mean-field
amplitude for an $N$-Boson system. In contrast to customary analysis, our
approach does not involve a splitting of the field operator into a condensate
part and a remainder. 
Instead, we will first introduce a candidate mean-field amplitude by taking
matrix elements of the field operator with the physical $N$-particle state
and certain auxiliary $(N-1)$-particle states, and then provide a criterion 
which decides whether or not that candidate is physically meaningful, so
that it qualifies as a macroscopic wave function.

Still staying within the framework of spontaneously broken symmetry, an 
appealing possibility to satisfy Eq.~(\ref{eq:SBS}) seems  to arise when 
$|\Psi\rangle$ is a {\em coherent state\/}, which, by definition, is an 
eigenstate of the annihilation operator:
\begin{equation}
	\ph_{\rm H}({\bm r};t_0) | \Psi \rangle = 
	\sqrt{N} \Phi({\bm r};t_0) | \Psi \rangle \; , 
\label{eq:CSI}
\end{equation}
or, upon return to the Schr\"odinger picture,
\begin{equation}
	\ph(\bm r) | \Psi(t_0) \rangle = 
	\sqrt{N} \Phi({\bm r};t_0) | \Psi(t_0) \rangle \; ;
\label{eq:TEE}
\end{equation}
in this respect such many-body coherent states are analogs of the familiar
standard harmonic-oscillator coherent states already introduced in the early 
days of quantum mechanics~\cite{Schrodinger26,Schiff68}. Intuitively speaking, 
Eq.~(\ref{eq:CSI}) means that $|\Psi\rangle$ is ``simple'' in the sense that 
it remains unchanged if one particle is removed. However, if $|\Psi\rangle$ 
is an eigenstate of $\ph_{\rm H}({\bm r};t_0)$ at one particular moment~$t_0$, 
it will not, in general, be one at others. Therefore, we are confronted with 
the necessity to admit more general concepts of ``coherence''. 

Following the traditional route for the time being and taking the 
symmetry-broken expectation value~(\ref{eq:SBS}), Eq.~(\ref{eq:HOM}) 
immediately yields 	 
\begin{eqnarray}
\label{eq:TRP}
	\ri \hbar \frac{\rd}{\rd t} \sqrt{N} \Phi\rt
	& = & h_1\rt \sqrt{N} \Phi\rt
\\	
	& + & g \left\langle \ph^\dagger_{\rm H}\rt 
	\ph_{\rm H}\rt \ph_{\rm H}\rt \right\rangle \; .
\nonumber
\end{eqnarray}
This leads to another pertinent problem: Of course one could write down the 
equation of motion for the third moment appearing here, but this would involve 
still higher moments, and so on, leading to an infinite chain of equations
expressing an infinite momentum hierarchy~\cite{KohlerBurnett02,ErdosEtAl06}. 
The Gross-Pitaevskii equation results from the assumption that this chain can 
be terminated at the lowest possible level, so that the expectation value of 
the triple operator product is replaced by the product of the expectation 
values of the individual operators: 
\begin{eqnarray}
	& & \left\langle \ph^\dagger_{\rm H}\rt 
	\ph_{\rm H}\rt \ph_{\rm H}\rt \right\rangle 
\nonumber \\	& \leadsto & 	
	\left\langle \ph^\dagger_{\rm H}\rt \right\rangle
	\left\langle \ph_{\rm H}\rt \right\rangle 
	\left\langle \ph_{\rm H}\rt \right\rangle \; .
\label{eq:TOP}
\end{eqnarray}
Here the ``$\leadsto$''-sign is meant to indicate that this closure 
hypothesis~(\ref{eq:TOP}) generally constitutes an uncontrolled approximation; 
note, however, that it would be exact for a coherent state~(\ref{eq:CSI}). 
Accepting this factorization, Eq.~(\ref{eq:TRP}) finally turns into the 
celebrated Gross-Pitaevskii equation~\cite{Gross61,Pitaevskii61,Gross63,
Gardiner97,CastinDum98,Leggett01,PethickSmith08,PitaevskiiStringari03,
ErdosEtAl06,ErdosEtAl07,ErdosEtAl09,ErdosEtAl10}, namely,     
\begin{equation}
	\ri \hbar \frac{\rd}{\rd t} \Phi\rt
	= h_1\rt \Phi\rt + Ng | \Phi\rt |^2 \Phi\rt \; .  
\label{eq:GPE}
\end{equation}
Since one has the identity
\begin{equation}
	\int \! \rd^3 r \left\langle \ph^\dagger_{\rm H}\rt
	\ph_{\rm H}\rt \right\rangle = N 
\label{eq:EVN}	
\end{equation}
for any $t$, one deduces from the definition~(\ref{eq:SBS}) that the  
mean-field amplitude is normalized to unity,
\begin{equation}
	\int \! \rd^3 r \, | \Phi\rt |^2 = 1 \; ,
\end{equation}
now assuming factorization of the expectation value~(\ref{eq:EVN}).

Besides providing an alternative definition of the mean-field amplitude not
based on spontaneous breach of symmetry, questioning and monitoring the 
accuracy of the required closure hypothesis constitutes a main objective 
of this work.  

We proceed as follows: In Sec.~\ref{sec:2} we outline a tentative construction 
process of a candidate mean-field amplitude associated with an $N$-particle 
state $| \Psi(t) \rangle_N$. This construction rests on the introduction of 
subsidiary $(N-1)$-particle states $| \wP(\bm r | t) \rangle_{N-1}$ which
are obtained from $| \Psi(t) \rangle_N$ by the annihilation of one particle;
the dynamics of these states are governed by the same Hamiltonian as those 
of the physical state $| \Psi(t) \rangle_N$. Our approach is similar in spirit 
to the introduction of a condensate wave function by Lifshitz and Pitaevskii 
in Ref.~\cite{LaLifIX}, but places particular emphasis on the way the 
initially ``neighboring'' states $| \Psi(t) \rangle_N$ and 
$| \wP(\bm r | t) \rangle_{N-1}$ evolve in time: Do they, in a suitable sense, 
remain close to each other? The subsequent derivation of the candidate's
equation of motion then makes clear under what conditions the Gross-Pitaevskii 
equation~(\ref{eq:GPE}) is satisfied, so that the construction yields a
{\em bona fide\/} macroscopic wave function. If, on the other hand, these 
conditions are not met, the candidate bears no physical meaning and has to be 
discarded.    
These considerations are not meant as a substitute for rigorous mathematical
analysis~\cite{ErdosEtAl06,ErdosEtAl07,ErdosEtAl09,ErdosEtAl10}, but rather 
as an elaboration of the salient features which make the Gross-Pitaevskii 
equation work. In particular, we isolate a notion of coherence which does 
not correspond, in the first place, to a single-particle orbital being 
macroscopically occupied, but which places a certain demand on the flow in 
Fock space instead, requiring its stiffness. 

The general scheme is applied in Sec.~\ref{sec:3} to a numerically tractable 
model system, describing a driven Josephson junction. The model is presented 
in Sec.~\ref{sub:31}, and the corresponding equations of motion are derived 
in Sec.~\ref{sub:32}, now paying particular attention to keeping all error 
terms. The results of selected numerical calculations which support our 
reasoning in considerable detail are discussed at length in Sec.~\ref{sub:33}. 
Besides being paradigmatically simple, the model has the further pleasant 
feature that it lends itself to a semiclassical interpretation, allowing one 
to relate the $N$-particle dynamics to those of a driven nonlinear pendulum. 
As laid out in Sec.~\ref{sub:34}, this semiclassical view leads to important 
qualitative insights which may guide the analysis of more complicated, 
experimentally accessible systems in the future. 
Thus, the model calculations collected in this Sec.~\ref{sec:3} complement
recent studies of periodically $\delta$-kicked Bose-Einstein condensates 
which aim at exploring nonlinear dynamics in a quantum many-body 
context~\cite{ZhangEtAl04,LiuEtAl06,DuffyEtAl04,WimbergerEtAl05,
BillamGardiner12,BillamEtAl13}; here we emphasize the potential usefulness 
of a sinusoidal force with a slowly varying envelope.  

We finally phrase our main findings, and draw some general conclusions, 
in Sec.~\ref{sec:4}.

\section{Constructive approach to the mean-field amplitude}
\label{sec:2}

We start by constructing the $N$-particle Fock-space state vector 
$| \Psi(t) \rangle_N$ corresponding to some given wave function $\Psi$: 
Building on a vacuum state $\vac$ characterized by 
\begin{equation}
	\langle {\rm vac} \vac = 1
\end{equation}	
and
\begin{equation}
	\ph(\bm r) \vac = 0 \quad \text{for each } \bm r \; , 
\label{eq:VAC}
\end{equation}
we have
\begin{eqnarray}
	| \Psi(t) \rangle_N & = & \frac{1}{\sqrt{N!}}
	\int \! \rd^3 r_1 \ldots \int \! \rd^3 r_N \,
	\Psi(\bm r_1, \ldots, \bm r_N; t)
\nonumber \\ & & \times \; 	
	\phd(\bm r_1) \ldots \phd(\bm r_N) \vac \; .
\label{eq:STV}
\end{eqnarray}
Repeatedly using the identity
\begin{eqnarray}
	& & \ph(\bm r') \prod_{j=1}^N \phd(\bm r_j) \vac
\nonumber \\	& = &	 
	\sum_{k=1}^N \delta(\bm r' - \bm r_k) 
	\prod_{j \ne k} \phd(\bm r_j) \vac  
\end{eqnarray}
which follows directly from Eqs.~(\ref{eq:BCR}) and (\ref{eq:VAC}), one 
confirms that this state is properly normalized, 
\begin{equation}
	{_N\langle} \Psi(t) | \Psi(t) \rangle_N = 1 \; , 	
\end{equation}
provided the Schr\"odinger wave function is,
\begin{equation}
	\int \! \rd^3 r_1 \ldots \int \! \rd^3 r_N \,
	| \Psi(\bm r_1, \ldots, \bm r_N; t) |^2 = 1 \; .
\end{equation}
This state vector~(\ref{eq:STV}) evolves in time according to the 
Schr\"odinger equation
\begin{equation}
	\ri\hbar\frac{\rd}{\rd t} | \Psi(t) \rangle_N
	= \Hh(t) | \Psi(t) \rangle_N
\end{equation}
with the many-body Hamiltonian~(\ref{eq:MBH}).

The system's one-particle reduced density matrix, as given by	
\begin{equation}
	\varrho(\bm r, \bm r'; t) = 
	{_N\langle} \Psi(t) | \phd(\bm r) \ph(\bm r') | \Psi(t) \rangle_N \; ,
\label{eq:OPR}
\end{equation}	
then takes the explicit form
\begin{eqnarray}
\label{eq:RDM}
	\varrho(\bm r, \bm r'; t) & = &	 
	N \int \! \rd^3 r_2 \ldots \int \! \rd^3 r_N \,
\\	& \times & 	
	\Psi^*(\bm r, \bm r_2, \ldots, \bm r_N; t)
	\Psi(\bm r', \bm r_2, \ldots, \bm r_N; t) \; .
\nonumber
\end{eqnarray}
It might appear natural to introduce a time-dependent macroscopic wave 
function on the basis of the Penrose-Onsager criterion for Bose-Einstein 
condensation in an interacting $N$-particle system~\cite{PenroseOnsager56}: 
At each moment~$t_0$ one can, in principle, perform a spectral decomposition 
\begin{equation}
	\varrho(\bm r, \bm r'; t_0) = 
	\sum_i n_i(t_0) \varphi_i^*(\bm r, t_0) \varphi_i(\bm r', t_0)	
\label{eq:SDD}
\end{equation}
of this density matrix~(\ref{eq:OPR}) into a complete orthonormal set of 
single-particle eigenfunctions $\varphi_i(\bm r, t_0)$. If then only one 
of the eigenvalues, say $n_0(t_0)$, is on the order of $N$, while all other 
ones are at most of order ${\mathcal O}(1)$, one has a simple Bose-Einstein 
condensate at $t_0$, and the eigenfunction $\varphi_0(\bm r, t_0)$ can be
regarded as the instantaneous macroscopic wave function. However, according to 
this construction the phase of $\varphi_0(\bm r, t_0)$ remains undetermined,
and unrelated to the phase of $\varphi_0(\bm r, t_1)$ at another moment~$t_1$,
whereas the Gross-Pitaevskii equation~(\ref{eq:GPE}) for $\Phi\rt$ constitutes
an initial-value problem which does not leave such a freedom. Hence, even in 
the most ideal case in which $n_0(t_0) = N$ for all $t_0$, it is not possible 
to reconstruct the full time-dependent macroscopic wave function from the 
one-particle reduced density matrix, even if one knows the exact many-body 
state at all times, because the phase of the macroscopic wave function 
evidently requires additional consideration. 

Therefore, here we take a different route. In situations where a mean-field 
amplitude $\Phi$ can be meaningfully defined, the most obvious requirement to 
be placed on such an object is that its absolute square, multiplied by $N$, 
be equal to the diagonal elements of the density matrix, that is, to the 
expectation values of the density operator, at least to a good approximation:    
\begin{equation}
	N | \Phi\rt |^2 = 
	{_N\langle} \Psi(t) | \phd(\bm r) \ph(\bm r) | \Psi(t) \rangle_N \; .
\label{eq:REQ}
\end{equation}	
Given an initial state $| \Psi(t_0)\rangle_N$ at some arbitrary initial 
moment~$t_0$, it is easy to satisfy this requirement on a formal level at 
that moment~$t_0$ by invoking the normalized $(N-1)$-particle states 
\begin{equation}
	| \wP(\bm r | t_0) \rangle_{N-1} := 
	\re^{\ri \gamma(\bm r)}
	\frac{\ph(\bm r) | \Psi(t_0)\rangle_N}
	     {\| \ph(\bm r) | \Psi(t_0)\rangle_N \|} \; , 
\label{eq:AUI}
\end{equation}
where $\gamma(\bm r)$ is a phase that can be suitably adjusted later on, 
and we have 
\begin{eqnarray}
	& & \| \ph(\bm r) | \Psi(t_0)\rangle_N \|^2 
\nonumber \\	& = &	
	N \int \! \rd^3 r_2 \ldots \int \! \rd^3 r_N \,
	| \Psi(\bm r, \bm r_2, \ldots, \bm r_N; t_0) |^2 
\end{eqnarray}		
according to Eq.~(\ref{eq:RDM}), as corresponding to the particle density at 
the position~$\bm r$. Thus, it may happen that the denominator appearing in 
the definition~(\ref{eq:AUI}) vanishes at certain points $\bm r$; such points 
either are exempted from the analysis, or require special treatment. 

It is quite essential not to be misled here: $| \wP(\bm r | t_0) \rangle_{N-1}$
must not be regarded as some sort of wave function depending on $\bm r$;
rather, for each $\bm r$ it is a state vector in the $(N-1)$-particle 
sector of the Fock space. We will mark such subsidiary $(N-1)$-particle states
by the ``tilde''-symbol in the following.

Introducing the projection operator
\begin{equation}
	\widehat{P}_{t_0} =
	| \wP(\bm r | t_0) \rangle_{N-1} \; {_{N-1}\langle} \wP(\bm r | t_0) |
	\; ,
\label{eq:PRI}	
\end{equation}	
one obviously has
\begin{eqnarray}
	& & 
	{_N\langle} \Psi(t_0) | \phd(\bm r) \ph(\bm r) | \Psi(t_0) \rangle_N
\nonumber \\	
	& = & {_N\langle} \Psi(t_0) | \phd(\bm r) \widehat{P}_{t_0} 
	\ph(\bm r) | \Psi(t_0) \rangle_N
\end{eqnarray}	
by the very definition~(\ref{eq:AUI}); as a consequence, we now may set
\begin{equation}		
	\sqrt{N} \Phi(\bm r,t_0) := 
	{_{N-1}\langle} \wP (\bm r|t_0) | \ph(\bm r) | \Psi(t_0) \rangle_{N}
	\; .
\label{eq:ITD}	
\end{equation}
Evidently the projector~(\ref{eq:PRI}) does not depend on the phase 
$\gamma(\bm r)$ of the subsidiary states, but this phase allows one to 
tune the phase of $\Phi(\bm r,t_0)$.

This tentative construction~(\ref{eq:ITD}) still does not mean that the
expression~$\Phi(\bm r,t_0)$ actually is a macroscopic wave function
at the moment~$t_0$: The above prescription could be applied to {\em any\/}
initial state $| \Psi(t_0) \rangle_N$, regardless of whether or not this state 
admits a mean-field description. At this point, a suitable counterpart of the 
Penrose-Onsager eigenvalue criterion which distinguishes a condensate is still 
lacking.

The usefulness of this construction becomes evident only when it is 
extended to later times: Besides the evolving physical $N$-particle state 
\begin{equation}
	| \Psi(t) \rangle_N = \Uh(t,t_0) \, | \Psi(t_0) \rangle_N
\label{eq:AES}
\end{equation}
we also have a family, parametrized by $\bm r$, of auxiliary $(N-1)$-particle
states which are obtained by subjecting the subsidiary initial 
states~(\ref{eq:AUI}) to the evolution generated by the very same Hamiltonian, 
giving
\begin{equation}
	| \wP(\bm r | t) \rangle_{N-1} = 
	\Uh(t,t_0) \, | \wP(\bm r | t_0) \rangle_{N-1} \; .
\label{eq:AUF}
\end{equation}
We now {\em define\/} a candidate mean-field amplitude $\Phi$ for all 
$t \ge t_0$ by taking, in literal extension of the initial-time 
definition~(\ref{eq:ITD}), the matrix elements of the field operator with
these evolving states~(\ref{eq:AES}) and (\ref{eq:AUF}): 
\begin{equation}		
	\sqrt{N} \Phi\rt := 
	{_{N-1}\langle} \wP (\bm r|t) | \ph(\bm r) | \Psi(t) \rangle_{N} \; .
\label{eq:DEF}
\end{equation}
Note that, in contrast to the ``symmetry-breaking'' definition~(\ref{eq:SBS}),
this still tentative definition is free of arbitrariness. The right-hand side 
of Eq.~(\ref{eq:DEF}) can in principle be evaluated once the initial state
$| \Psi(t_0) \rangle_N$ is given; there is no freedom comparable to the
choice of the coefficients $a_N$ in Eq.~(\ref{eq:SUP}). However, it is not 
yet clear at this stage whether the quantity $\Phi$ defined in this manner 
actually conforms to the central requirement~(\ref{eq:REQ}) for $t > t_0$, 
and whether or not it satisfies the Gross-Pitaevskii equation. Obviously, the 
answers to these questions depend to a large extent on $| \Psi(t_0) \rangle_N$,
but also on the particular time-dependence imposed on the Hamiltonian. On the 
other hand, it is easy to see that our construction does fulfill its purpose 
in the ideal case where all $N$ particles initially occupy the same 
orbital~$\varphi$: Then the initial many-body wave function is given by a 
Hartree product state,  
\begin{equation}
	\Psi_\varphi(\bm r_1, \ldots , \bm r_N; t_0) = 
	\prod_{j=1}^N \varphi(\bm r_j; t_0) \; ;
\label{eq:PWF}
\end{equation}
in this case one expects $\Phi(\bm r, t_0)$ to be equal to 
$ \varphi(\bm r,t_0)$.

Indeed, following the general prescription~(\ref{eq:STV}) the state 
$|\Psi_\varphi (t_0) \rangle_{N}$ associated in Fock space with this 
product wave function~(\ref{eq:PWF}) acquires the form 
\begin{equation}
	| \Psi_\varphi (t_0) \rangle_{N} = \frac{1}{\sqrt{N!}}
	\left[ 
	\int \! \rd^3 r \, \varphi(\bm r; t_0) \, \phd(\bm r) 
	\right]^N \vac 
\label{eq:GCS}
\end{equation}
and thus obeys the quasi-eigenvalue equation 
\begin{equation}
	\ph(\bm r) | \Psi_\varphi (t_0) \rangle_{N} = 
	\sqrt{N} \varphi(\bm r;t_0) | \Psi_\varphi (t_0) \rangle_{N-1} \; ,  	
\label{eq:QEE}
\end{equation}
implying that the one-particle reduced density matrix~(\ref{eq:OPR}) here 
becomes
\begin{equation}
	\varrho(\bm r, \bm r'; t_0) =
	N \varphi^*(\bm r; t_0) \varphi(\bm r'; t_0) \; .
\label{eq:DCS}
\end{equation}	
This reflects precisely the Penrose-Onsager criterion for Bose-Einstein 
condensation~\cite{PenroseOnsager56}: In the case of a pure condensate the 
one-particle reduced density matrix~(\ref{eq:SDD}) equals the $N$-fold of the 
projection operator onto the $N$-fold occupied single-particle orbital. 
Moreover, Eq.~(\ref{eq:QEE}) immediately yields
\begin{equation}
	\frac{\ph(\bm r) | \Psi_\varphi (t_0) \rangle_{N}}
	     {\| \ph(\bm r) | \Psi_\varphi (t_0) \rangle_{N} \|}
	= \frac{\varphi(\bm r;t_0)}{| \varphi(\bm r;t_0) |} \,	      
	| \Psi_\varphi (t_0) \rangle_{N-1} \; .  	
\end{equation}  
Hence, if one chooses the phase $\gamma(\bm r)$ as the negative phase of 
the single-particle orbital $\varphi(\bm r, t_0)$, so that, for instance,
$\gamma(\bm r) \equiv 0$ when $\varphi(\bm r, t_0)$ is real and positive, 
Eq.~(\ref{eq:AUI}) here reduces to
\begin{equation}
	| \wP_\varphi(\bm r | t_0) \rangle_{N-1} =
	| \Psi_\varphi( t_0) \rangle_{N-1} \; .
\label{eq:COS}	
\end{equation}
In conjunction with the quasi-eigenvalue equation~(\ref{eq:QEE}) and our 
definition~(\ref{eq:DEF}) of the mean-field amplitude, this furnishes the 
expected initial identity
\begin{equation}
	\Phi(\bm r, t_0) = \varphi(\bm r, t_0) \; .
\label{eq:EQP}
\end{equation}	
What may be even more important, Eq.~(\ref{eq:COS}) also drastically simplifies
the entire construction process for {\em all\/} subsequent times $t > t_0$, 
because the family~(\ref{eq:AUF}) reduces to a single member not depending 
on $\bm r$, and for all times the dependence of $\Phi$ on $\bm r$ in 
Eq.~(\ref{eq:DEF}) is inflicted solely by the field operator $\ph(\bm r)$. 

The Hartree product wave function~(\ref{eq:PWF}) embodies the idea of
``simplicity'' in a particularly straightforward manner. If the many-body
wave function carries merely the information content of a single-particle 
orbital, no information is lost when employing that orbital as a mean-field
amplitude, in the spirit of a macroscopic wave function. For the sake of easy 
nomenclature we will refer to the associated Fock-space states~(\ref{eq:GCS})
as $N$-particle coherent --- in short: $N$-{\em coherent\/} --- in the
following. This wording appears to be somewhat at odds with the 
quasi-eigenvalue equation~(\ref{eq:QEE}), which is reminiscent of the 
action of an annihilation operator on an harmonic-oscillator eigenstate, 
or on a typical Fock state, in marked distinction to the proper eigenvalue 
equation~(\ref{eq:TEE}) characterizing a conventional coherent state. But
it is this type of coherence which is implied by macroscopic occupation with 
a sharp number of particles. Needless to say, an initially $N$-coherent state 
generally will not remain $N$-coherent in the course of time. Thus, the 
question what property an $N$-particle state must have in order to admit a 
Gross-Pitaevskii description of its time evolution still needs to be settled.

Returning to the general case not necessarily starting from an $N$-coherent
state~(\ref{eq:GCS}), the equation of motion of our candidate mean-field 
amplitude~(\ref{eq:DEF}) is given by 
\begin{equation}
	\ri \hbar \frac{\rd}{\rd t} \sqrt{N} \Phi\rt =
	{_{N-1}\langle} \wP(\bm r|t) | \left[ \ph(\bm r) , \Hh(t) \right] |
	\Psi(t) \rangle_N \; .
\label{eq:MOC}
\end{equation}
In formal agreement with the previous Heisenberg equation~(\ref{eq:WOC}), the 
commutator yields
\begin{eqnarray}
\label{eq:COM} 	
	\left[ \ph(\bm r) , \Hh(t) \right] & = & h_1\rt \ph(\bm r)
\\	& + &
	\int \! \rd^3 r' \, 
	\phd(\bm r') U(\bm r, \bm r') \ph(\bm r') \ph(\bm r) \; ,
\nonumber
\end{eqnarray}		
having exploited the symmetry~(\ref{eq:TPI}). Now the first term on the 
right-hand side leads to the matrix elements 
\begin{eqnarray}
	& & {_{N-1}\langle} \wP(\bm r|t) | h_1\rt \ph(\bm r) |
	\Psi(t) \rangle_N 
\nonumber \\	& \leadsto &
	h_1\rt \; {_{N-1}\langle} \wP(\bm r|t) | \ph(\bm r) |
	\Psi(t) \rangle_N
\nonumber \\	& = &
	h_1\rt \sqrt{N} \Phi\rt \; ,	
\label{eq:WFT}
\end{eqnarray}
where, in analogy to Eq.~(\ref{eq:TOP}), 
the ``$\leadsto$''-symbol indicates that this particular step can be correct 
for some special states, but will lead to uncontrolled errors for most others. 
That is, steps marked by this symbol determine about the quality of our 
construction: The magnitude of the error committed there decides whether or
not it will result in an acceptable mean-field amplitude.
In the present case~(\ref{eq:WFT}) the single-particle Hamiltonian $h_1$, when 
willfully drawn before the matrix element in order to recover the candidate 
mean field amplitude~(\ref{eq:DEF}), also acts on the $\bm r$-dependence of 
the auxiliary family~(\ref{eq:AUF}) --- unless there is no such dependence. 
At this point, the observation made below Eq.~(\ref{eq:EQP}), deriving from 
the property~(\ref{eq:COS}) of $N$-coherent states, leads to a noteworthy 
insight:  
\begin{quote}
	{\em If the initially given $N$-particle state is exactly $N$-coherent, 
	in the sense specified by Eqs.~(\ref{eq:PWF}) and (\ref{eq:GCS}), 
	it suffices to employ the corresponding $(N-1)$-particle 
	state~(\ref{eq:COS}) as sole subsidiary state, implying that the 
	chain~(\ref{eq:WFT}) is an equality for all later times $t > t_0$.} 
\end{quote}

Thus, one may reasonably hope that the error admitted when working with 
close-to-$N$-coherent initial states remains tolerably small, but substantial 
further work seems required to turn this vague hope into a quantitative 
statement.           

The second term on the right-hand side of Eq.~(\ref{eq:COM}) confronts us with
a different problem: Acting twice with the field operator on an $N$-particle
state would bring us to the $(N-2)$-particle sector of Fock space, over which
we have no command. Therefore, with the help of the commutation 
relations~(\ref{eq:BCR}) we reorder the operators such that their successive
application produces images which alternate between the $(N-1)$- and the
$N$-particle sector:
\begin{eqnarray}
	& & \int \! \rd^3 r' \, 
	\phd(\bm r') U(\bm r, \bm r') \ph(\bm r') \ph(\bm r)  
\nonumber \\	& = &
	\int \! \rd^3 r' \, U(\bm r, \bm r')
	\ph(\bm r) \phd(\bm r') \ph(\bm r') 
\nonumber \\	& - &
	\int \! \rd^3 r' \, U(\bm r, \bm r')
	\delta (\bm r - \bm r') \ph(\bm r') \; . 			
\label{eq:REO}
\end{eqnarray}
The first contribution appearing here is dealt with by replacing the actual
interaction potential by the effective contact interaction~(\ref{eq:COP}),
giving the triple operator product $g \ph(\bm r) \phd(\bm r)\ph(\bm r)$.
Taking the matrix elements required by the definition~(\ref{eq:DEF}), we 
then write   
\begin{eqnarray}
	& & g \; {_{N-1}\langle} \wP(\bm r|t) | \ph(\bm r) \phd(\bm r) 
	\ph(\bm r) | \Psi(t) \rangle_N
\nonumber \\	& \leadsto &
	g \; {_{N-1}\langle} \wP(\bm r|t) | \ph(\bm r) | \Psi(t) \rangle_N 
\nonumber \\	& \times & \qquad  		
	{_{N}\langle} \Psi(t) | \phd(\bm r) | \wP(\bm r | t) \rangle_{N-1} 
\nonumber \\	& \times & \quad 	
	{_{N-1}\langle} \wP(\bm r|t) | \ph(\bm r) | \Psi(t) \rangle_N
\nonumber \\	& = & g N^{3/2} | \Phi\rt |^2 \Phi\rt \; . 		
\label{eq:SFC}
\end{eqnarray}	
Note that ``$\leadsto$'' appears once again: Here the projection operators 
projecting onto the states $| \wP(\bm r | t) \rangle_{N-1}$ and 
$| \Psi(t) \rangle_N$, respectively, have been inserted between the 
field operators ``as suitable'', but so far without deeper justification, 
thus providing a formal substitute for the standard closure 
assumption~(\ref{eq:TOP}). We will investigate the validity of this step 
exemplarily in Sec.~\ref{sec:3}.

The second contribution arising from Eq.~(\ref{eq:REO}) necessitates to 
evaluate the interaction potential $U(\bm r, \bm r')$ at zero distance,
$\bm r = \bm r'$, and thus forbids us to employ the effective 
substitute~(\ref{eq:COP}). This appears to be not more than a spurious 
complication, so that here we restrict ourselves to interaction potentials 
which remain finite at zero distance, such as a simple repulsive step 
potential~\cite{WeissEtAl05}. Then $U(\bm r, \bm r) \sim c g$ will be about 
proportional to the coupling constant~$g$, with a factor~$c$ carrying the
dimension of an inverse volume. Thus, after taking the required matrix
elements we are left with 
\begin{equation}
	c g \; 
	{_{N-1}\langle} \wP(\bm r|t) | \ph(\bm r) | \Psi(t) \rangle_N
	= c g \sqrt{N} \Phi\rt \; .
\label{eq:SPP}
\end{equation}
Collecting the elements~(\ref{eq:WFT}), (\ref{eq:SFC}), and (\ref{eq:SPP}),
the basic equation of motion~(\ref{eq:MOC}) adopts the form
\begin{eqnarray}
	\ri \hbar \frac{\rd}{\rd t} \Phi\rt & = & h_1\rt \Phi\rt
\nonumber \\	& + &	 
	Ng \left( | \Phi\rt |^2 - \frac{c}{N} \right) \Phi\rt \; .   
\end{eqnarray}
Taking the limits $N \to \infty$ and $g \to 0$ such that the product 
$Ng$ remains constant, this actually becomes the Gross-Pitaevskii 
equation~(\ref{eq:GPE}), confirming that our definition~(\ref{eq:DEF}) 
indeed has the potential to work as desired, still with the proviso that both 
the initial state $| \Psi(t_0) \rangle_N$ and the Hamiltonian $\Hh(t)$ be such 
that the replacements indicated by ``$\leadsto$'' in Eqs.~(\ref{eq:WFT}) and 
(\ref{eq:SFC}) constitute at least good approximations.  
 
However, to rederive the Gross-Pitaevskii equation was not what we wanted to 
achieve in the first place. Rather, we had set out to construct a mean-field
amplitude $\Phi$ such that its absolute square yields the exact expectation
values of the density operator according to Eq.~(\ref{eq:REQ}). Is this the 
case now? 

Evidently, in order to factorize the right-hand side of Eq.~(\ref{eq:REQ})
exactly we should be writing    
\begin{eqnarray}
	& & 
	{_N\langle} \Psi(t) | \phd(\bm r) \ph(\bm r) | \Psi(t) \rangle_N
\nonumber \\	& = &
	{_N\langle} \Psi(t) | \phd(\bm r) \widehat{Q}_t \ph(\bm r) | 
	\Psi(t) \rangle_N		 
\label{eq:GFE}
\end{eqnarray}	
with projection operators 
\begin{equation}
	\widehat{Q}_t = 
	| \widetilde\Xi(\bm r| t) \rangle_{N-1} \; 
	{_{N-1}\langle} \widetilde\Xi(\bm r| t) |
\label{eq:PRE}
\end{equation}
constructed from the normalized states
\begin{eqnarray}
 	| \widetilde\Xi(\bm r | t) \rangle_{N-1} & = &
	\frac{\ph(\bm r) | \Psi(t) \rangle_N}
	     {\| \ph(\bm r) | \Psi(t) \rangle_N \|}
\nonumber \\	& = &
	\frac{\ph(\bm r) \Uh(t,t_0) | \Psi(t_0) \rangle_N}
	     {\| \ph(\bm r) | \Psi(t) \rangle_N \|} 		     
\label{eq:EFA}
\end{eqnarray}
which are obtained by first propagating $| \Psi(t_0) \rangle_N$ in time,
and then annihilating a particle at $\bm r$. Instead, Eq.~(\ref{eq:REQ})
combined with the definition~(\ref{eq:DEF}) implies that the success of
our construction would be guaranteed if one had the identity
\begin{eqnarray}
	& & 
	{_N\langle} \Psi(t) | \phd(\bm r) \ph(\bm r) | \Psi(t) \rangle_N
\nonumber \\	& = &
	{_N\langle} \Psi(t) | \phd(\bm r) \widehat{P}_t \ph(\bm r) | 
	\Psi(t) \rangle_N \; ,
\label{eq:HAI}		 
\end{eqnarray}	
where, in continuation of Eq.~(\ref{eq:PRI}), the operator 
\begin{equation}
	\widehat{P}_t =
	| \wP(\bm r | t) \rangle_{N-1} \; {_{N-1}\langle} \wP(\bm r | t) |
\end{equation}	
projects onto the ray generated by
\begin{equation}
	\wP(\bm r | t) \rangle_{N-1} = 
	\re^{\ri\gamma(\bm r)}
	\frac{\Uh(t, t_0) \ph(\bm r) | \Psi(t_0)\rangle_N}
	     {\| \ph(\bm r) | \Psi(t_0)\rangle_N \|} \; . 
\label{eq:SUC}
\end{equation}
In contrast to the previous state~(\ref{eq:EFA}), this is the state found by 
first removing a particle from $| \Psi(t_0) \rangle_N$ at $\bm r$, and then 
propagating from $t_0$ to $t$. These two projection operators $\widehat{P}_t$ 
and $\widehat{Q}_t$ are identical, meaning that Eq.~(\ref{eq:HAI}) actually 
is an exact equality, {\em if\/} $| \widetilde\Xi(\bm r | t) \rangle_{N-1}$ 
differs from $| \wP(\bm r | t) \rangle_{N-1}$ merely by a phase factor, and 
thus is proportional to $| \wP(\bm r | t) \rangle_{N-1}$. This latter demand 
requires 
\begin{equation}
	\ph(\bm r) | \Psi(t) \rangle_N =
	\eta\rt \, | \wP(\bm r | t) \rangle_{N-1}  
\end{equation}
with some function $\eta\rt$; taking the scalar product with 
$| \wP(\bm r | t) \rangle_{N-1}$ immediately yields
\begin{equation}
	\eta\rt = \sqrt{N} \Phi\rt \; .
\end{equation}	
The resulting consistency condition
\begin{eqnarray}
	& & 
	\ph(\bm r) \Uh(t,t_0) | \Psi(t_0) \rangle_N 
\nonumber \\	& = &
	\sqrt{N} \Phi\rt \re^{\ri \gamma(\bm r)}
	\frac{\Uh(t,t_0) \ph(\bm r) | \Psi(t_0)\rangle_N}
	     {\| \ph(\bm r) | \Psi(t_0)\rangle_N \|} \; ,   
\label{eq:COC}
\end{eqnarray}
or, in more compact form,
\begin{equation}
	\ph(\bm r) | \Psi(t) \rangle_N =
	\sqrt{N} \Phi\rt | \wP(\bm r | t) \rangle_{N-1} \; ,
\label{eq:TCO}
\end{equation}
constitutes an interesting generalization of the quasi-eigenvalue 
equation~(\ref{eq:QEE}). Whereas that former relation characterizes
$N$-particle coherence at one fixed moment~$t_0$ only, its 
descendant~(\ref{eq:TCO}) intrinsically incorporates the system's time 
evolution: Annihilating a particle from the time-evolved state should 
produce the state obtained by annihilating prior to evolution; 
states having this property will be called {\em $t$-coherent\/} from here 
onwards.

Thus, we may phrase the preceding considerations as follows:    
\begin{quote}
	{\em If the $N$-particle state under consideration is $t$-coherent, 
	in the sense specified by Eqs.~(\ref{eq:COC}) or (\ref{eq:TCO}), 
	the absolute square of the mean-field amplitude defined by 
	Eq.~(\ref{eq:DEF}) equals the particles' density, obeying 
	Eq.~(\ref{eq:REQ}).\/}   
\end{quote}

Obviously, the desired identity~(\ref{eq:TCO}) selects those cases in which 
a good macroscopic wave function does exist. Even if it is satisfied initially 
to a good approximation, it may cease to be valid under the action of external 
forcing. The value of the above conceptions now rests in the recognition that 
they furnish a tool for monitoring the achievable performance of the 
Gross-Pitaevskii equation: What really matters is the {\em degree of 
$t$-coherence\/}, as expressed by the extent to which the states~(\ref{eq:EFA})
required for exact factorization equal the states~(\ref{eq:SUC}) underlying 
our construction; this extent is quantified by the absolute value of the 
``direction cosine''~\cite{GertjerenkenHolthaus15}
\begin{eqnarray}
\label{eq:DCO}
	R\rt & = & \re^{\ri\gamma(\bm r)}{_{N-1}\langle} \wP (\bm r | t) |
	\widetilde{\Xi}(\bm r | t) \rangle_{N-1}
\\		& = &
	\re^{\ri\gamma(\bm r)}
	\frac{{_{N-1}\langle} \wP (\bm r | t) | \ph(\bm r) 
	       | \Psi(t) \rangle_N}
	     {\| \ph(\bm r) | \Psi(t) \rangle_{N} \|}
\nonumber \\	& = &
	\frac{{_N\langle} \Psi(t_0) | \phd(\bm r) \widehat{U}^\dagger(t,t_0) 
	        \ph(\bm r) \Uh(t,t_0) | \Psi(t_0) \rangle_N}
	     {\| \ph(\bm r) | \Psi(t_0) \rangle_N \| \; 
	      \| \ph(\bm r) | \Psi(t) \rangle_N \| } \; .
\nonumber	      	     	
\end{eqnarray}
Namely, if $| R\rt | = 1$ the solution $\Phi\rt$ to the Gross-Pitaevskii
equation does provide an accurate description of the system; if 
$0 \le | R\rt | < 1$, it is bound to miss some aspects of the full
$N$-particle dynamics. In that sense, $R\rt$ quantifies the ``degree of 
mean-field approximability'' of the evolving system at time~$t$, and at the 
position~$\bm r$. Dynamically speaking, the magnitude $| R\rt |$ evaluates
the evolution of two trajectories in Fock space initially differing by one 
particle; if it does not matter whether one annihilates a particle  first and 
then lets the system evolve over some interval of time, or evolves first and 
annihilates then, the flow in Fock space may be considered {\em stiff\/}. 
Therefore, we refer to the magnitude $| R\rt |$ as {\em stiffness\/}, with 
maximum stiffness $| R\rt | = 1$ indicating perfect $t$-coherence. Note that
such maximum stiffness still does not mean that annihilation and evolution
commute in the operator-theoretic sense, but rather that the system manages 
to hide the non-commutativity of these operations for the state considered 
behind a phase factor which drops out of the projectors involved.

It is also of interest to observe that if Eq.~(\ref{eq:TCO}) is satisfied
the second line of Eq.~(\ref{eq:DCO}) yields
\begin{equation}
	R\rt = \re^{\ri\gamma(\bm r)} \frac{\Phi\rt}{| \Phi\rt |} \; ,	
\end{equation}
so that the phase of the solution to the Gross-Pitaevskii equation, 
which is relatively easy to obtain numerically, contains information 
about how $|\widetilde{\Xi}(\bm r | t) \rangle_{N-1}$ differs from   
$| \wP(\bm r | t) \rangle_{N-1}$, while both of these states designate
the same ray in Fock space;
this information quantifies the difference between the evolution of an $N$-- 
and that of an $(N-1)$-particle state. On the one hand, this observation 
ties up with the well-known fact that in thermodynamic equilibrium, when the 
Hamiltonian is time-independent, the phase of the macroscopic wave function 
is given by the chemical potential, {\em i.e.\/}, by the energy required to 
add one more particle to the system. On the other hand, it also clarifies why 
it is not possible to determine the phase of the macroscopic wave function
from the one-particle reduced density matrix~(\ref{eq:SDD}), even if the
Penrose-Onsager criterion is optimally satisfied: That matrix draws its 
information solely from the $N$-particle sector of Fock space. 
 
It goes without saying that the theoretical construction process outlined
in the present section is not meant as a practical help for solving 
$N$-particle evolution equations; rather, it provides a conceptual framework
within which time-dependent many-Boson systems can be analyzed. The following 
study of a model which allows one to carry out all of the above steps at least 
numerically will confirm that the general terms coined here do indeed capture 
essential features of $N$-Boson dynamics in a nontrivial manner.

\section{Case study: The driven two-mode system}
\label{sec:3}

\subsection{The model}
\label{sub:31}

We now investigate in detail a system of $N$ identical spinless repulsively 
interacting Bose particles which are confined to two sites, labeled $1$ and 
$2$. Each pair of Bosons occupying the same site is assumed to invariably 
contribute the repulsion energy $2\hbar\kappa$ to the total energy of the 
system, while intersite interaction is neglected. Moreover, there is a 
tunneling contact between the two sites, with hopping matrix element 
$\hbar\Omega/2$, so that $\Omega$ denotes the single-particle tunneling 
frequency. In terms of the operators $\ajd$ and $\aj$ which create and 
annihilate, respectively, a particle at the $j$th site, obeying the Bose 
commutation relations
\begin{equation}
	\left[ \aj, a_k \right] = 0 		\; , \quad 
	\left[ \ajd, a_k^\dagger \right] = 0 	\; , \quad
	\left[ \aj, a_k^\dagger \right] = \delta_{jk} 
\end{equation}	
with $j,k = 1,2$, the system then is described by the Josephson 
Hamiltonian~\cite{Leggett01}
\begin{equation}
	H_0 = -\frac{\hbar\Omega}{2}\left( \aLd\aR + \aRd\aL\right)
	+ \hbar\kappa\left(\aLd\aLd\aL\aL + \aRd\aRd\aR\aR\right) \; . 
\label{eq:LMG}
\end{equation}
This may be regarded as a particular variant of a model introduced by Lipkin, 
Meshkov, and Glick in order to test the validity of many-body approximation 
methods~\cite{LipkinEtAl65,MeshkovEtAl65,GlickEtAl65}, in a spirit quite 
similar to that of the present paper. It is realized approximately by 
a Bose-Einstein condensate loaded into an optical double-well potential, 
thus providing a ``bosonic Josephson junction''~\cite{GatiMKO07}; 
in this context the model has met with substantial renewed 
interest~\cite{MilburnEtAl97,ParkinsWalls98,MahmudEtAl05,BoukobzaEtAl09,
JuliaDiazEtAl10,NissenKeeling10,ChuchemEtAl10,SimonStrunz12,GraefeEtAl14}.
For ease of notation we write its Fock-space operators without ``hat''-symbol. 

Obviously the general field operator $\ph(\bm r)$ with its continuous 
``index'' $\bm r$, as considered in the previous section, here corresponds to 
the pair $a_1$, $a_2$. Likewise, the macroscopic wave function $\Phi\rt$ now 
reduces to a set of two discrete mean-field amplitudes $c_1(t)$, $c_2(t)$,
the absolute squares of which, multiplied by $N$, should give the numbers 
of particles residing at the two sites. Introducing the dimensionless time 
variable $\tau = \Omega t$, their equations of motion, easily derived by 
the scheme sketched in the Introduction, take the form     
\begin{eqnarray}
	\ri\frac{\rd}{\rd \tau} c_1(\tau) & = & -\frac{1}{2} c_2(\tau)
	+ 2\alpha | c_1(\tau) |^2 c_1(\tau) 
\nonumber \\	
	\ri\frac{\rd}{\rd \tau} c_2(\tau) & = & -\frac{1}{2} c_1(\tau)
	+ 2\alpha | c_2(\tau) |^2 c_2(\tau) \; , 
\label{eq:GPM}
\end{eqnarray} 
depending only on the scaled interaction strength 
\begin{equation}
	\alpha = \frac{N\kappa}{\Omega} \; .
\label{eq:SIS}
\end{equation}	
This system~(\ref{eq:GPM}) constitutes the Gross-Pitaevskii equation for 
the Lipkin-Meshkov-Glick model~(\ref{eq:LMG}); it also describes 
self-trapping on a polaronic dimer~\cite{EilbeckEtAl85,KenkreCampbell86}. 
It is well known that this system can be integrated ana\-lytically in terms 
of Jacobian elliptic functions~\cite{KenkreCampbell86,RaghavanEtAl99}. 
But such integrability clearly signals that the model~(\ref{eq:LMG}) still 
is too simple for our present purposes: For testing the viability of a
Gross-Pitaevskii-type approximation to full $N$-particle evolution under
typical conditions we require a system which exhibits more generic, 
non-integrable mean-field dynamics. For this reason we extend the model by 
adding an explicitly time-dependent driving term: Respecting the general form 
of a ``canonical Josephson Hamiltonian''~\cite{Leggett01}, we study the driven 
two-mode system   
\begin{equation}
	H(t) = H_0 + H_1(t) \; ,
\label{eq:HDJ}
\end{equation}
where
\begin{equation}
	H_1(t) = \hbar\mu(t)\sin(\omega t)\left(\aLd\aL - \aRd\aR \right)
\label{eq:HDR}
\end{equation}
models a bias applied with frequency $\omega$ between the two sites, while 
its amplitude $\hbar\mu(t)$ may vary in time. Even with fixed amplitude 
$\mu(t) = \mu_1$ this driven bosonic Josephson junction possesses chaotic 
mean-field solutions~\cite{HolthausStenholm01,WeissTeichmann08,
WeissTeichmann09,GertjerenkenHolthaus14}.

Adapting the construction~(\ref{eq:STV}), a general $N$-particle state of
the model~(\ref{eq:HDJ}) can be written in the form
\begin{equation}
	| \Psi(t) \rangle_N = \frac{1}{\sqrt{N!}}
	\sum_{j_1=1,2} \!\! \ldots \!\! \sum_{j_N=1,2} 
	\psi_{j_1,\ldots,j_N}\!(t)
	a_{j_1}^\dagger \ldots 	a_{j_N}^\dagger \vac \; .
\end{equation}
Because of bosonic symmetry the coefficients 
$\psi_{j_1,\ldots,j_N}\!(t)$ coincide for all permutations of their indices.
Writing $\psi_{j_1,\ldots,j_N}\!(t) = \psi_n(t)$ when $n$ of the $N$ indices
$j_1, \ldots, j_N$ equal $1$ (so that the other $N-n$ indices equal $2$),
we then have
\begin{eqnarray}
	| \Psi(t) \rangle_N & = & \frac{1}{\sqrt{N!}}
	\sum_{n=0}^N \binom{N}{n} \psi_n(t) 
	\left( \aLd \right)^n \left( \aRd \right)^{N-n} \vac
\nonumber \\	& = &
	\sum_{n=0}^N \binom{N}{n}^{1/2} \psi_n(t) \, | n, N-n \rangle \; ,
\end{eqnarray}
where
\begin{equation}
	| n, N-n \rangle = \frac{\left( \aLd \right)^n}{\sqrt{n!}}
	\frac{\left( \aRd \right)^{N-n}}{\sqrt{(N-n)!}} \, \vac
\label{eq:FOS}
\end{equation}
denotes the Fock state with $n$ particles on site~$1$ and, consequently,
$N-n$ particles on site~$2$. Thus, a general $N$-particle state is specified 
by $N+1$ time-dependent complex coefficients $\psi_n(t)$.  

An $N$-particle coherent state which corresponds, in the sense of 
Eq.~(\ref{eq:GCS}), to an $N$-fold occupied single-particle orbital is 
characterized by the requirement that each creation operator be accompanied by 
the same amplitude. Denoting the single-particle basis of the two-site system 
as $\{ \varphi_1, \varphi_2 \}$, and parametrizing a general single-particle 
orbital $\varphi$ in this basis by means of two angles $\theta$, $\phi$ in 
the form   
\begin{equation}
	\varphi = \cos(\theta/2) \varphi_1 
	+ \sin(\theta/2) \re^{\ri \phi} \varphi_2 \; , 
\label{eq:NSP}
\end{equation}
thus omitting an irrelevant overall phase, the coherent-state coefficients 
become 
\begin{equation}
	\psi_n = \cos^n(\theta/2) \sin^{N-n}(\theta/2) \re^{\ri(N-n)\phi} \; ;
\end{equation}
the $N$-particle coherent states of the two-site model are therefore 
written as
\begin{eqnarray} 
	| \theta,\phi \rangle_N & = &
	\sum_{n=0}^N \binom{N}{n}^{1/2} \cos^n(\theta/2)
\nonumber \\	& & \times	
	\sin^{N-n}(\theta/2) \re^{\ri(N-n)\phi} | n, N-n \rangle \; .
\label{eq:ACS}
\end{eqnarray}
These states, which emerge here as special cases of the $N$-coherent 
states~(\ref{eq:GCS}), play a significant role in quantum optics, in which 
field they are referred to as ``$SU(2)$-coherent states'' or ``atomic coherent 
states''~\cite{ArecchiEtAl72}. In the present context their most important 
property lies in the fact that they satisfy the quasi-eigenvalue equations 
\begin{eqnarray}
	\aL | \theta,\phi \rangle_N & = & 
	\;\;\;\, \sqrt{N} \cos(\theta/2) \, | \theta,\phi \rangle_{N-1}
\nonumber \\
	\aR | \theta,\phi \rangle_N & = & 
	\sqrt{N} \sin(\theta/2) \re^{\ri\phi} \, | \theta,\phi \rangle_{N-1} \; ,
\label{eq:TMQ}
\end{eqnarray}	
in formal correspondence with the general Eq.~(\ref{eq:QEE}).

\begin{figure}[tb]
\begin{center}
\includegraphics[width = 0.9\linewidth]{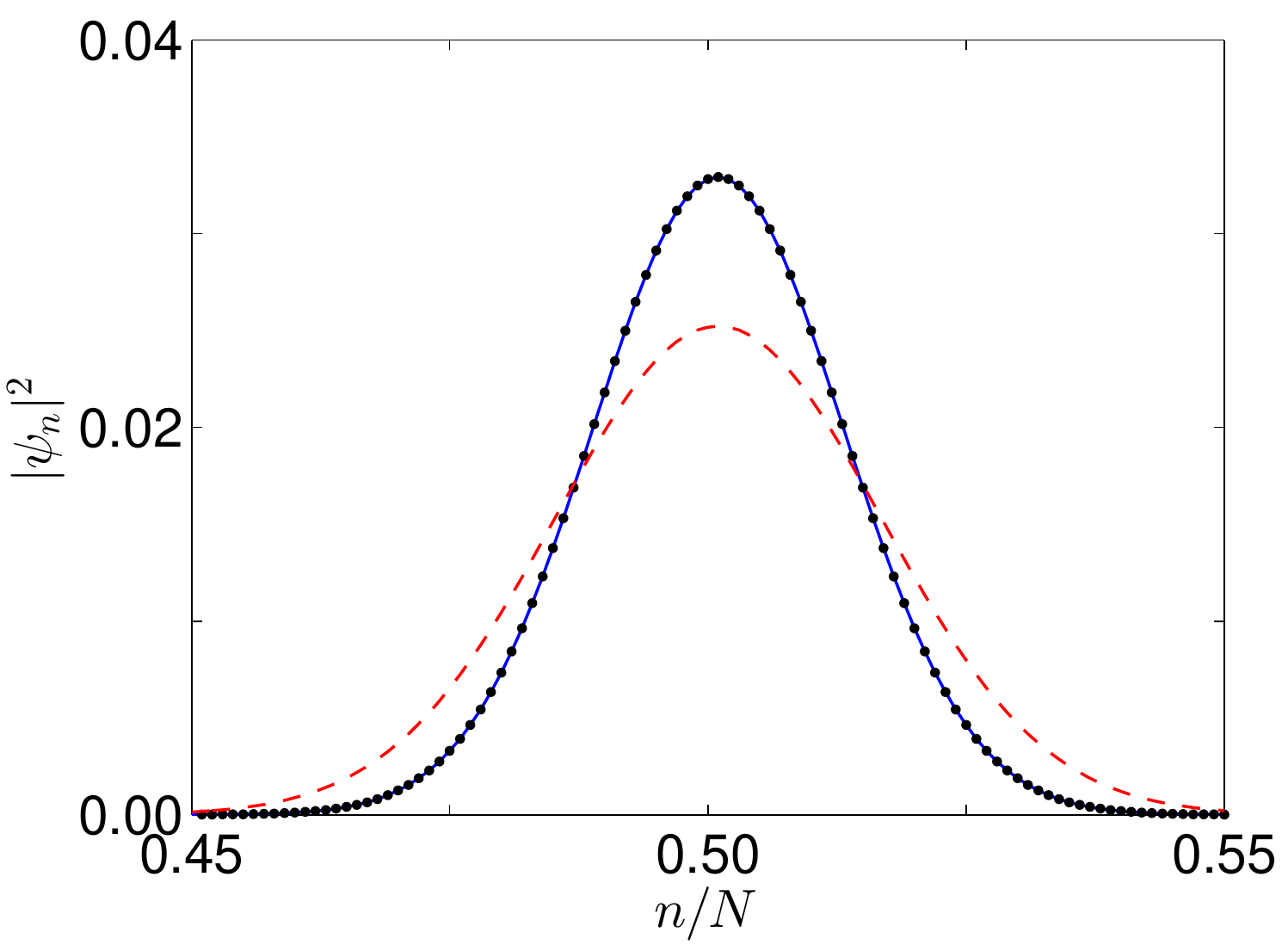}
\end{center}
\caption{(Color online) Comparison of the exact ground state of the two-mode
	model~(\ref{eq:LMG}) for $\alpha = 0.95$ and $N = 1000$ (dots) with 
	the corresponding variational state~(\ref{eq:VAR}) with optimal width 
	parameter~(\ref{eq:OPT}) (full line), and with the $N$-coherent 
	state~(\ref{eq:SCS}) (dashed). Plotted are the squares of the
	respective expansion coefficients $\psi_n$ in the Fock-state basis.}   
\label{F_1}
\end{figure}

It is easy to see that the best $N$-coherent approximation to the ground
state of the undriven Josephson junction~(\ref{eq:LMG}) is provided by
the symmetric state   
\begin{equation}
	| \pi/2,0 \rangle_N = \frac{1}{2^{N/2}}
	\sum_{n=0}^N \binom{N}{n}^{1/2} | n, N-n \rangle \; , 
\label{eq:SCS}	
\end{equation}	
giving
\begin{equation}
	\frac{{_N\langle} \pi/2,0 | H_0 | \pi/2,0 \rangle_N}{N\hbar\Omega}
	= -\frac{1}{2} + \frac{\alpha}{2}\left( 1 - \frac{1}{N} \right)		
\end{equation}
with the scaled interaction strength~(\ref{eq:SIS}). However, this 
$N$-coherent state~(\ref{eq:SCS}) is an eigenstate of the 
$\hbar\Omega$-proportional ``hopping'' term of the Hamiltonian~(\ref{eq:LMG}),
but not of its $\hbar\kappa$-proportional interaction term, which would be
diagonalized by the Fock states~(\ref{eq:FOS}) instead. Therefore, the
$N$-coherent state~(\ref{eq:SCS}) is a good approximation to the true
ground state only as long as $\alpha \ll 1$; for stronger interaction 
the true ground state must be more sharply peaked around $n = N/2$. 
Because one has 
\begin{equation}
	\binom{N}{n}^{1/2} \approx \binom{N}{N/2}^{1/2} 
	\exp\Big(-\frac{1}{N} (n - N/2)^2 \Big)
\label{eq:AFB}
\end{equation}
for $|n - N/2| \ll N/2$, a good ansatz for the exact ground state is  
\begin{equation}
	| \gamma \rangle_N = \left( \frac{2\gamma}{\pi} \right)^{1/4}
	\sum_{n=0}^N \exp\Big( -\gamma(n - N/2)^2 \Big) |n, N-n \rangle
\label{eq:VAR}
\end{equation}
with variational parameter $\gamma$, supposing $N \gg 1$. An elementary 
calculation then yields
\begin{eqnarray}
	\frac{{_N\langle} \gamma | H_0 | \gamma \rangle_N}{N\hbar\Omega}
	& = & -\frac{1}{2}
	\left(1 + \frac{1}{N} - \frac{\gamma}{2} -\frac{1}{2\gamma N^2}\right)
\nonumber \\ 	& &
	+ \frac{\alpha}{2}\left(1 - \frac{2}{N} + \frac{1}{\gamma N^2} \right)
\nonumber \\ 	& &	
	+ \mathcal{O}(N^{-2}) \; ,
	\phantom{\frac{1}{2}} 		
\end{eqnarray}
where the neglected terms of order $\mathcal{O}(N^{-2})$ do not depend on
$\gamma$. Hence, one deduces the optimal parameter
\begin{equation}
	\gamma_{\rm opt} = \frac{\sqrt{2\alpha + 1}}{N}
\label{eq:OPT}
\end{equation}	
which properly reduces to $\gamma_{\rm opt} \approx 1/N$ when $\alpha \ll 1$,
as expected from the approximation~(\ref{eq:AFB}), but grows with increasing
interaction strength so as to suppress fluctuations. Figure~\ref{F_1} compares 
this variational ground state for $\alpha = 0.95$ and $N = 1000$ to the 
numerically computed exact one, confirming the quality of the variational 
approach. More generally, one has  
\begin{equation}
	\big| {_N\langle} \pi/2,0 | \gamma_{\rm opt} \rangle_N \big|^2
	= \frac{2(2\alpha + 1)^{1/4}}{\sqrt{2\alpha + 1} + 1} 
\end{equation}
for large~$N$, implying that the ground state of the Josephson 
junction~(\ref{eq:LMG}) is quite different from the $N$-coherent 
state~(\ref{eq:SCS}) for large $\alpha$, and does not approach it in the 
limit $N \to \infty$ when this limit is taken such that $\alpha$ remains
constant. This will become important for assessing the numerical studies 
reported in Sec.~\ref{sub:33}.

\subsection{Equations of motion}
\label{sub:32}

The goal now is to explicitly carry through the construction process outlined 
in Sec.~\ref{sec:2}, to monitor its accuracy, and to keep track of the errors 
committed, for the driven two-mode system~(\ref{eq:HDJ}). We are given the 
Schr\"odinger equation  
\begin{equation}
	\ri\hbar \frac{\rd}{\rd t} | \Psi(t) \rangle =
	H(t) | \Psi(t) \rangle
\label{eq:TMS}
\end{equation}
for $N$-particle Fock-space states $| \Psi(t) \rangle$ which are written 
without particle number index here, and we wish to construct mean-field 
amplitudes $c_j(t)$ such that  
\begin{equation}
	N | c_j(t) |^2 = \langle \Psi(t) | \ajd \aj | \Psi(t) \rangle
\label{eq:TBR}
\end{equation}
for $j = 1,2$, in compliance with the central requirement~(\ref{eq:REQ}). 
Starting from an initial $N$-particle state $| \Psi(t_0) \rangle$ prepared 
at time $t = t_0$, we follow the prescription~(\ref{eq:AUI}) and construct 
the two subsidiary $(N-1)$-particle states 
\begin{equation}
	| \wP_j(t_0) \rangle := \re^{\ri \gamma_j}
	\frac{\aj | \Psi(t_0) \rangle}{\| \aj | \Psi(t_0) \rangle \|} \; ,
\label{eq:TSS}
\end{equation}
where the phases $\gamma_j$ can be chosen at will, and we are assuming
that both sites are initially occupied, so that the denominators do not
vanish. These states obviously permit the factorizations
\begin{eqnarray} & & 
	\langle \Psi(t_0) | \ajd \aj | \Psi(t_0) \rangle
\nonumber \\	& = &	
	\langle \Psi(t_0) | \ajd | \wP_j(t_0) \rangle
	\langle \wP_j(t_0) | \aj | \Psi(t_0) \rangle	
\end{eqnarray}
and therefore allow one to define the candidate mean-field amplitudes at the
initial moment: 
\begin{equation}
	\sqrt{N} c_j(t_0) := 
	\langle \wP_j(t_0) | \aj | \Psi(t_0) \rangle \; .
\end{equation}
Then not only the actual initial state $| \Psi(t_0) \rangle$ is subjected 
to the time evolution generated by $H(t)$, as expressed by the original 
Schr\"odinger equation~(\ref{eq:TMS}), but also the two subsidiary
states~(\ref{eq:TSS}), giving rise to two further evolution equations 
\begin{equation}
	\ri\hbar \frac{\rd}{\rd t} | \wP_j(t) \rangle =
	H(t) | \wP_j(t) \rangle \; .
\label{eq:SGS}
\end{equation}
Here a decisive feature of the Fock-space formalism is exploited: Regardless 
of whether it acts on the $N$-particle sector or on the $(N-1)$-particle 
sector of Fock space, the Hamiltonian $H(t)$ is the same. Taking the solutions 
to these equations~(\ref{eq:TMS}) and (\ref{eq:SGS}), the candidate mean-field 
amplitudes then are defined for all times $t \ge t_0$ in accordance with 
Eq.~(\ref{eq:DEF}):   
\begin{equation}
	\sqrt{N} c_j(t) := 
	\langle \wP_j(t) | \aj | \Psi(t) \rangle \; .
\label{eq:TDF}
\end{equation}
Their equations of motion are obtained from the identity
\begin{equation}
	\ri\hbar \frac{\rd}{\rd t}
	\langle \wP_j(t) | \aj | \Psi(t) \rangle = 
	\langle \wP_j(t) | \, \left[ \aj, H(t) \right] \, | 
	\Psi(t) \rangle \; .
\end{equation}	
Evaluating the commutator
\begin{equation} 
	\left[\aL, H(t) \right] = -\frac{\hbar\Omega}{2} \aR
	+ 2\hbar\kappa \aLd\aL\aL + \hbar\mu(t)\sin(\omega t) \aL 
\label{eq:TMC}	
\end{equation}
and taking the required matrix elements, one is led to 
\begin{eqnarray}
	\ri\hbar \frac{\rd}{\rd t} \sqrt{N} c_1(t) & = &
	-\frac{\hbar\Omega}{2}\sqrt{N} c_2(t) f_1(t)
\nonumber \\	& &		
	+ 2\hbar\kappa \langle \wP_1(t) | \aLd \aL \aL |
	\Psi(t) \rangle
	\phantom{\frac{\hbar\Omega}{2}}
\nonumber \\	& &	
	+ \hbar\mu(t) \sin(\omega t) \sqrt{N} c_1(t) \; ,  	
\label{eq:TME}
\end{eqnarray}
where we have introduced the quantity 
\begin{equation}
	f_1(t) = \frac{\langle \wP_1(t) | \aR | \Psi(t) \rangle}
	              {\langle \wP_2(t) | \aR | \Psi(t) \rangle}
	\; ;		      
\label{eq:RAT}
\end{equation}
an analogous, still exact equation holds for $c_2(t)$. Observe the role 
of the ratio~(\ref{eq:RAT}): When operating on Eq.~(\ref{eq:TMC}) with  
$|\wP_1(t)\rangle$ from the left, and with $| \Psi(t) \rangle$ from the right, 
the tunneling term proportional to $\aR$ yields its numerator, whereas the 
desired amplitude $c_2(t)$ is given by its denominator. Wilfully setting 
$f_1(t) \equiv 1$, and doing the same with the ratio $f_2(t)$ appearing in 
the equation for $c_2(t)$, corresponds to the uncontrolled step marked by 
``$\leadsto$'' in the general Eq.~(\ref{eq:WFT}); retaining both $f_1(t)$ and 
$f_2(t)$ in Eq.~(\ref{eq:TME}) and its $c_2(t)$-counterpart therefore renders 
these equations exact, and allows one to control one type of error accepted in 
the usual Gross-Pitaevskii approach. In the special case that the initial state 
$| \Psi (t_0) \rangle$ is given by an atomic coherent state~(\ref{eq:ACS}),
the quasi-eigenvalue equations~(\ref{eq:TMQ}) ensure that 
$|\wP_1(t_0)\rangle = |\wP_2(t_0)\rangle = |\theta,\phi\rangle_{N-1}$,
having set $\gamma_1 = 0$, $\gamma_2 = -\phi$, and hence one deduces
$|\wP_1(t)\rangle = |\wP_2(t)\rangle$ for {\em all\/} times~$t \ge t_0$. 
Therefore, in this special case one actually has $f_1(t) = f_2(t) \equiv 1$, 
so that the above error does not occur. However, in all other cases it needs 
to be considered.  

To proceed with Eq.~(\ref{eq:TME}), we now adapt the general steps 
(\ref{eq:REO}) and (\ref{eq:SFC}) for processing the triple operator
products:
\begin{eqnarray} & & 		      
	\langle \wP_j(t) | \ajd \aj \aj | \Psi(t) \rangle
\nonumber \\	& = &	
	\langle \wP_j(t) | \big( \aj \ajd - 1 \big) \aj |
	\Psi(t) \rangle
\nonumber \\	& = &
	\langle \wP_j(t) | \aj \ajd \aj | \Psi(t) \rangle 
	- \sqrt{N} c_j(t)
\nonumber \\	& \leadsto &
	\langle \wP_j(t) | \aj | \Psi(t) \rangle 	
	\langle \Psi(t) | \ajd | \wP_j(t) \rangle 
	\langle \wP_j(t) | \aj | \Psi(t) \rangle 
\nonumber \\	& & 
	- \sqrt{N} c_j(t)
\nonumber \\	& = &
	N^{3/2} c_j(t) \left( | c_j(t) |^2 - \frac{1}{N} \right) \; .			 	
\end{eqnarray}
The error introduced when assuming the factorization following the
``$\leadsto$''-symbol is given by the difference 
\begin{equation}
	\Delta_j(t) = 
	\langle \wP_j(t) | \aj \ajd \aj | \Psi(t) \rangle
	- N^{3/2} | c_j(t) |^2 c_j(t) \; .
\label{eq:EOC}
\end{equation}	
Hence, we have the {\em exact\/} evolution equation
\begin{eqnarray}
	\frac{\ri}{\Omega} \frac{\rd}{\rd t} c_1(t) & = & 
	-\frac{1}{2}c_2(t) f_1(t) 
\nonumber \\	& & 		
	+ 2\frac{N\kappa}{\Omega} \left[ 
	\left( | c_1(t) |^2 - \frac{1}{N} \right) c_1(t)
	+ \frac{\Delta_1(t)}{N^{3/2}} \right]
\nonumber \\	& & 
	+ \frac{\mu(t)}{\Omega} \sin(\omega t) c_1(t) \; , 
\end{eqnarray}	
together with its counterpart for $c_2(t)$. Finally, taking the limit
$N \to \infty$ such that
\begin{equation}
	\frac{N\kappa}{\Omega} = \alpha = {\rm const.}
\label{eq:ALC}
\end{equation}
we obtain the Gross-Pitaevskii equation for the driven two-mode 
system~(\ref{eq:HDJ}), with the proviso that both
\begin{equation}
	f_j(t)	\longrightarrow 1
\label{eq:TLF}	
\end{equation}
and
\begin{equation}
	\frac{\Delta_j(t)}{N^{3/2}} \longrightarrow 0
\label{eq:TLD}	
\end{equation}
in this limit, for $j = 1,2$: Again employing the dimensionless time
variable $\tau = \Omega t$, we then have	
\begin{eqnarray}
	\ri \frac{\rd}{\rd \tau} c_1(\tau) & = & -\frac{1}{2} c_2(\tau)
	+ 2\alpha | c_1(\tau) |^2 c_1(\tau)
\nonumber \\	& + &	
	\frac{\mu(\tau)}{\Omega}\sin\left(\frac{\omega}{\Omega}\tau\right)
	c_1(\tau) \; ,
\nonumber \\
	\ri \frac{\rd}{\rd \tau} c_2(\tau) & = & -\frac{1}{2} c_1(\tau)
	+ 2\alpha | c_2(\tau) |^2 c_2(\tau)
\nonumber \\	& - &
	\frac{\mu(\tau)}{\Omega}\sin\left(\frac{\omega}{\Omega}\tau\right)
	c_2(\tau) \; .		
\label{eq:TSY}
\end{eqnarray}		
The key question, of course, is under what conditions the 
limits~(\ref{eq:TLF}) and~(\ref{eq:TLD}) actually are adopted, and how
relevant these limits are when~$N$ is still finite, and kept fixed. 

A further question deriving from the discussion in Sec.~\ref{sec:2} is to 
what extent the solution to this system~(\ref{eq:TSY}) does comply with the
basic requirement~(\ref{eq:TBR}). In analogy with Eqs.~(\ref{eq:GFE}) and
(\ref{eq:PRE}) we now have the exact identities
\begin{eqnarray} & & 
	\langle \Psi(t) | \ajd \aj | \Psi(t) \rangle
\nonumber \\	& = &	
	\langle \Psi(t) | \ajd | \widetilde{\Xi}_j(t) \rangle
	\langle \widetilde{\Xi}_j(t) | \aj | \Psi(t) \rangle	
\label{eq:TMG}
\end{eqnarray}
with
\begin{eqnarray}
	| \widetilde{\Xi}_j(t) \rangle & = &
	\frac{\aj | \Psi(t) \rangle}{\| \aj | \Psi(t) \rangle \|}
\nonumber \\	& = &
	\frac{ \aj  U(t,t_0) | \Psi(t_0) \rangle}{\| \aj | \Psi(t) \rangle \|}
	\; , 	
\end{eqnarray}
whereas we require 
\begin{eqnarray} & & 
	\langle \Psi(t) | \ajd \aj | \Psi(t) \rangle
\nonumber \\	& = &	
	\langle \Psi(t) | \ajd | \wP_j(t) \rangle
	\langle \wP_j(t) | \aj | \Psi(t) \rangle	
\label{eq:TMD}
\end{eqnarray}
with
\begin{equation}
	| \wP_j(t) \rangle = \re^{\ri\gamma_j}\frac{
	U(t,t_0) \aj | \Psi(t_0) \rangle}{ \| \aj | \Psi(t_0) \rangle \| }
	\; .
\end{equation}
The actually fulfilled Eq.~(\ref{eq:TMG}) is compatible with the desired
Eq.~(\ref{eq:TMD}) if $| \widetilde{\Xi}_j(t) \rangle$ differs from
$| \wP_j(t) \rangle$ merely by a phase factor, demanding a relation
\begin{equation}
	\aj | \Psi(t) \rangle = \eta_j(t) | \wP_j(t) \rangle
	\; ;
\end{equation}	
the definition~(\ref{eq:TDF}) of the candidate mean-field amplitude then 
immediately yields
\begin{equation}
	\eta_j(t) = \sqrt{N} c_j(t) \; .
\end{equation}
Thus, we are led to a set of two consistency equations analogous to
Eq.~(\ref{eq:TCO}) which embody the property of perfect $t$-coherence: 
\begin{equation}
	\aj | \Psi(t) \rangle = \sqrt{N} c_j(t) | \wP_j(t) \rangle \; ;
\label{eq:CON}
\end{equation}	
this is the property which guarantees the validity of the basic
Eq.~(\ref{eq:TBR}) underlying the entire construction process. 

Following Eq.~(\ref{eq:DCO}), the degree to which $t$-coherence actually does 
prevail, that is, the stiffness of Fock-space flow in the vicinity of the
initial state, is quantified by the scalar products 
\begin{eqnarray}
	R_j(t) & = & \re^{\ri \gamma_j}	
	\langle \wP_j(t) | \widetilde{\Xi}_j(t) \rangle
\nonumber \\	& = & \re^{\ri \gamma_j}
	\frac{\langle \wP_j(t) | \aj | \Psi(t) \rangle}
	{\| \aj | \Psi(t) \rangle \| }		   
\nonumber \\	& = &
	\frac{ \langle \Psi(t_0) | \ajd U^\dagger(t,t_0) \aj U(t,t_0)
	| \Psi(t_0) \rangle} 
	{ \| \aj | \Psi(t_0) \rangle \| \; \| \aj | \Psi(t) \rangle \|} \; .
\label{eq:TSP}
\end{eqnarray}
If both $| \widetilde{\Xi}_j(t) \rangle$ and $| \wP_j(t) \rangle$ 
represent the same ray in Fock space, as is implied by the consistency 
condition~(\ref{eq:CON}), then $| R_j(t)| = 1$; monitoring the absolute
value of this scalar product~(\ref{eq:TSP}) therefore allows one to assess, 
on the basis of the exact $N$-particle solutions $| \Psi(t) \rangle$ and
the $(N-1)$-particle solutions $| \wP_j(t) \rangle$, the possible accuracy
of a Gross-Pitaevskii-treatment. Moreover, if close-to-perfect $t$-coherence
is given and Eq.~(\ref{eq:CON}) is satisfied more or less exactly, then
one has
\begin{equation}
	R_j(t) = \re^{\ri \gamma_j} \frac{c_j(t)}{| c_j(t) |} \; ,
\label{eq:PHD}
\end{equation} 
so that the phase of $R_j(t)$ equals the phase of the respective mean-field
amplitude, up to $\gamma_j$. Read in the reverse direction, this means that 
if a Gross-Pitaevskii approach is viable, then the phase of the easily 
obtainable mean-field amplitudes contains valuable information about the exact 
many-particle dynamics, more precisely, on the difference between the evolution 
of $N$ and $(N-1)$ particles.

\subsection{Numerical studies}
\label{sub:33}

Since the dynamics of the driven Josephson junction~(\ref{eq:HDJ}), when
occupied with~$N$ particles, takes place in a merely $(N+1)$-dimensional 
complex Hilbert space, solving the equations of motion for moderately 
large~$N$ does not pose severe computational difficulties. For ensuring high
numerical accuracy we employ a variable-order Adams PECE algorithm, of the 
type originally elaborated by Shampine and Gordon~\cite{ShampineGordon75}, 
routinely reaching particle numbers~$N$ on the order of a few thousand.

\begin{figure}[t]
\begin{center}
\includegraphics[width = 0.9\linewidth]{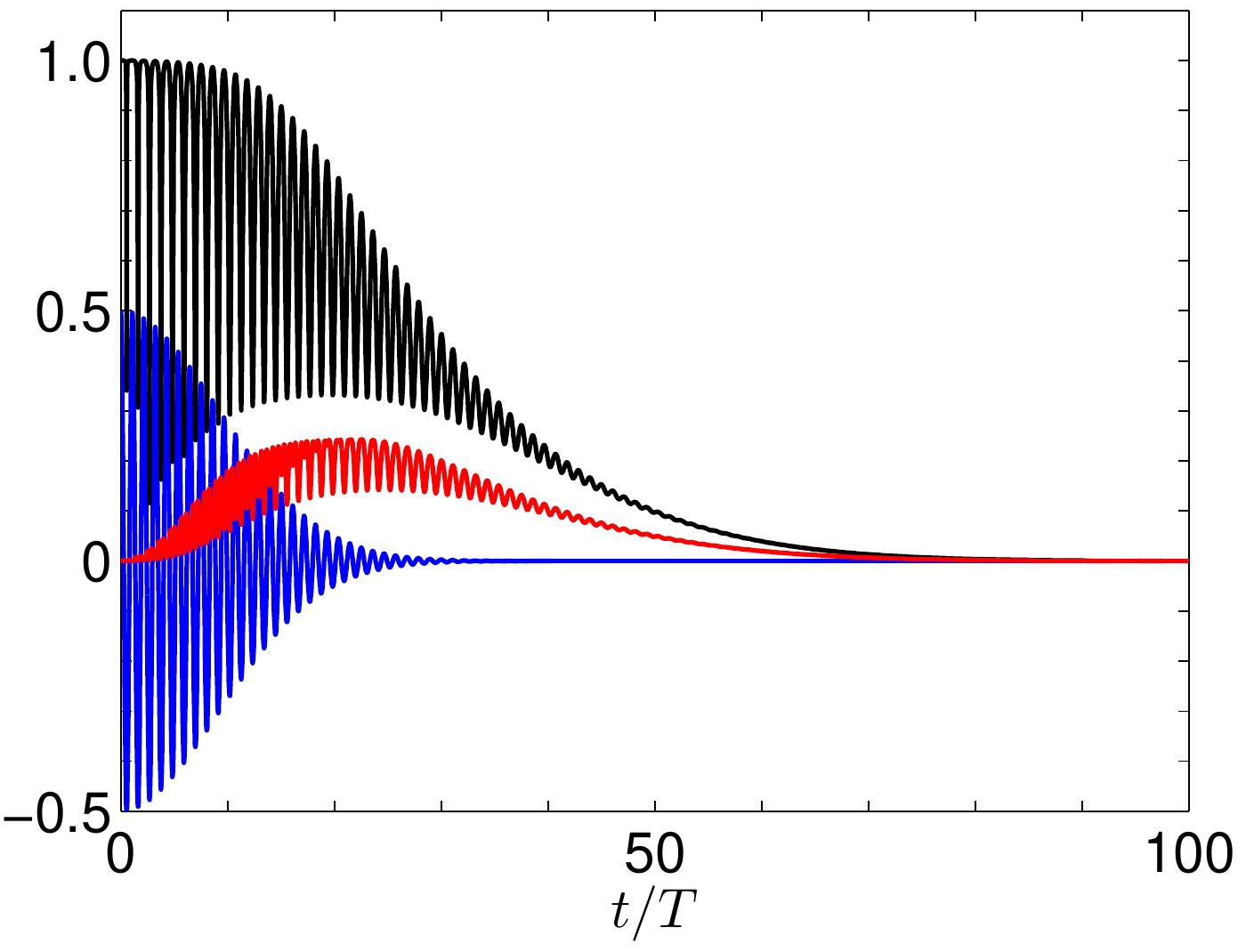}
\end{center}
\caption{(Color online) Scaled population imbalance $\langle J_z \rangle/N$ 
	(blue line), closure error $|\Delta_1|/N^{3/2}$ (red line), and 
	stiffness $|R_1|$ (black line) for the undriven Josephson 
	junction~(\ref{eq:LMG}) with $\alpha = 0.5$ and $N = 1000$. The 
	initial state is the Fock state $|N,0\rangle$, which is $N$-coherent. 
	Here and in the following figures, time~$t$ is measured in multiples 
	of $T = 2\pi/\Omega$.}
\label{F_2}
\end{figure}

\begin{figure}[th]
\begin{center}
\includegraphics[width = 0.9\linewidth]{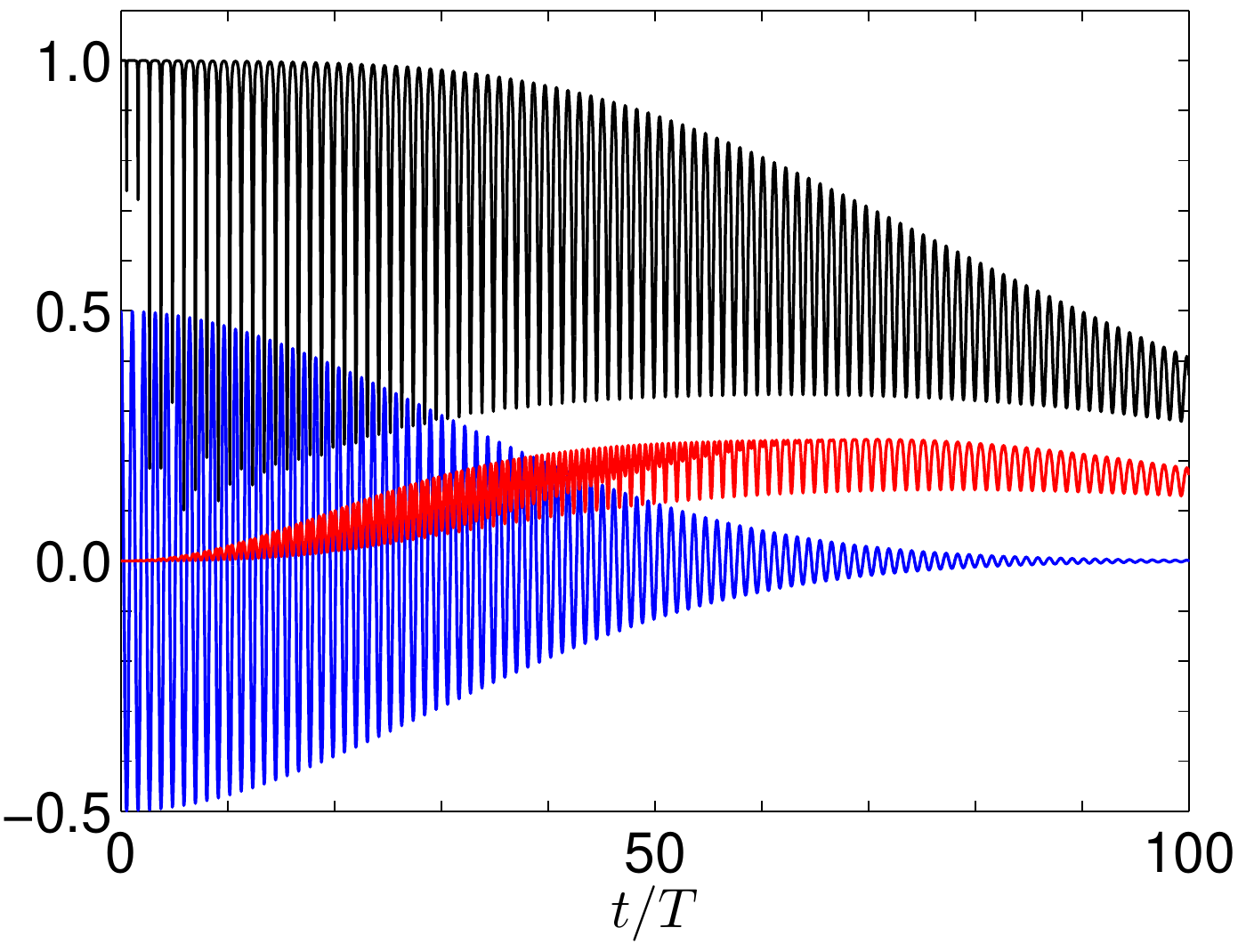}
\end{center}
\caption{(Color online) As Fig.~\ref{F_2}, but with $N = 10000$.}
\label{F_3}
\end{figure}

The diagnostic tools now at our disposal are the ratios~(\ref{eq:RAT}),
the errors of closure~(\ref{eq:EOC}), and the stiffnesses~(\ref{eq:TSP}).
While the first and the second of these tools allow one to trace sources 
of deviations from Gross-Pitaevskii behavior, the third one evaluates the
degree of $t$-coherence of the evolving many-body wave function, admitting
a truthful Gross-Pitaevskii description only when $|R_j(t)| = 1$.   

\begin{figure}[t]
\begin{center}
\includegraphics[width = 0.9\linewidth]{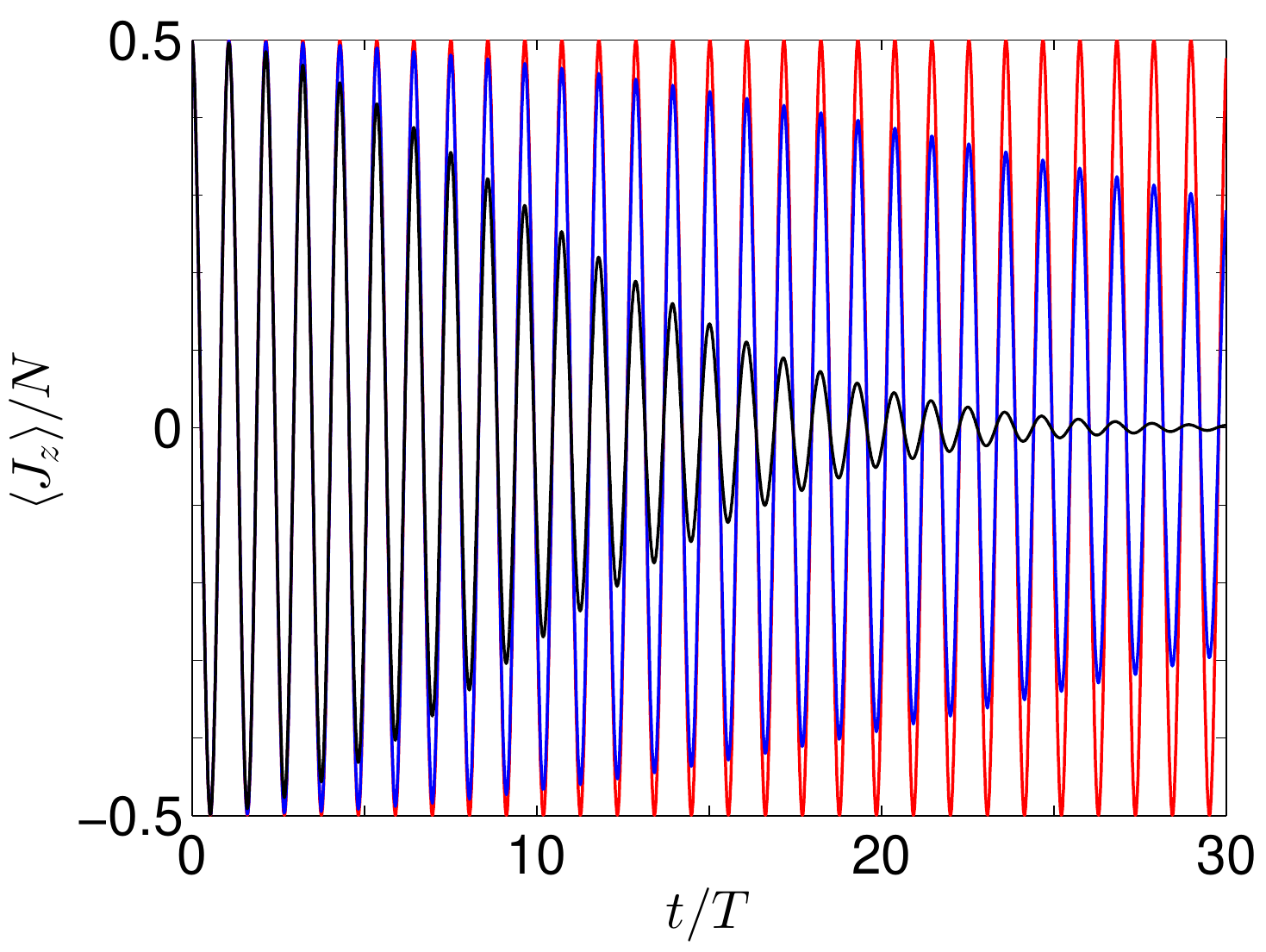}
\end{center}
\caption{(Color online) Scaled population imbalance $\langle J_z \rangle/N$
	for the undriven Josephson junction~(\ref{eq:LMG}) with $\alpha = 0.5$
	for $N=1000$ (black line) and $N=10000$ (blue line), in comparison
	with the prediction of the Gross-Pitaevskii equation~(\ref{eq:GPM})
	(red line). In each case, all particles are initially located at 
	site~1.}   
\label{F_4}
\end{figure}

As preliminary examples for the performance of these tools we monitor 
the dynamics of the junction~(\ref{eq:LMG}) in the absence of the 
drive~(\ref{eq:HDR}). As initial condition we select 
$|\Psi(0)\rangle = |N, 0\rangle$, 
so that all particles occupy site~1 at $t_0 = 0$. Then 
$|\wP_1(0)\rangle = |N-1,0\rangle$ 
is given by Eq.~(\ref{eq:TSS}), with $\gamma_1 = 0$. Moreover, since the 
initial Fock state $|N,0\rangle$ equals an $N$-coherent state~(\ref{eq:ACS}) 
with $\theta = 0$ and arbitrary $\phi$, we may set 
$|\wP_2(t)\rangle = |\wP_1(t)\rangle$ for $t \ge 0$. 
This implies $f_1(t) = f_2(t) \equiv 1$, so that all deviations from 
mean-field dynamics are solely due to the closure errors. Figure~\ref{F_2} 
depicts the scaled population imbalance
\begin{equation}
	\langle J_z \rangle(t)/N = 
	\langle \Psi(t) | \aLd \aL - \aRd \aR | \Psi(t) \rangle / (2N)
\label{eq:INP}
\end{equation}
for $N = 1000$ and scaled interaction strength $\alpha = 0.5$, together with
the scaled error of closure $|\Delta_1(t)|/N^{3/2}$ and the stiffness
$|R_1(t)|$, with time~$t$ being measured in multiples of $T = 2\pi/\Omega$.
The quantities $|\Delta_2(t)|/N^{3/2}$ and $|R_2(t)|$ behave quite similar
to their respective counterparts. One observes the familiar collapse of the 
oscillating population imbalance, caused by dephasing due to the finite 
particle number, which is to be followed by revivals at later times, 
when the components of the wave function rephase~\cite{SimonStrunz12,
HolthausStenholm01,WrightEtAl96,LewensteinYou96}. The most significant error 
of closure occurs during the collapse stage, leading to a stiffness which 
decreases to zero in an oscillating manner. When increasing the particle 
number to $N = 10000$, while keeping $\alpha = 0.5$ fixed so that, in 
accordance with Eq.~(\ref{eq:ALC}), the interparticle interaction strength 
$\hbar\kappa$ is reduced by a factor of $10$, the collapse proceeds more 
slowly, as plotted in Fig.~\ref{F_3}. The comparison of the $N$-particle 
dynamics for both $N = 1000$ and $N = 10000$ with the prediction 
\begin{equation}
	\langle J_z \rangle_{\rm mf}(t)
	= \Big( |c_1(t) |^2 - | c_2(t) |^2 \Big) / 2
\label{eq:IMF}
\end{equation}
of the Gross-Piaevskii equation~(\ref{eq:GPM}) displayed in Fig.~\ref{F_4} 
provides convincing evidence for convergence in the limit $N \to \infty$, 
when $\alpha$ is held constant: Given any moment $t_1 > 0$, the particle 
number~$N$ can be increased such that the true populations of the two sites, 
calculated from the $N$-particle Schr\"odinger equation, differ by an 
arbitrary small amount from the corresponding mean-field data in the entire 
interval from $t_0 = 0$ to $t_1$ in the example considered here. But as 
pointed out in Sec.~\ref{sub:31}, this situation is exceptionally simple 
insofar as the mean-field equation of motion~(\ref{eq:GPM}) is integrable.

For assessing the conceptually far more difficult, but generic cases of
non-integrable mean-field dynamics we now turn to the driven Josephson
junction~(\ref{eq:HDJ}), and choose a Gaussian envelope function  
\begin{equation}
	\mu(t) = \mu_{\rm max} \exp\left(-\frac{t^2}{2\sigma^2}\right) \; ,
\label{eq:GAP} 
\end{equation}
so that the drive~(\ref{eq:HDR}) models a single pulse with maximum
amplitude~$\hbar\mu_{\rm max}$, carrier frequency~$\omega$, and width~$\sigma$.
For all following calculations we set $\omega/\Omega = 1.0$ and 
$\sigma/T = 10.0$, where again $T = 2\pi/\Omega$. Thus, the driving frequency
is at resonance with the single-particle tunneling frequency, and the
driving amplitude increases smoothly from almost zero to its maximum within 
about $30$~driving cycles, then decreasing within another $30$~cycles.

\begin{figure}[b]
\begin{center}
\includegraphics[width = 0.9\linewidth]{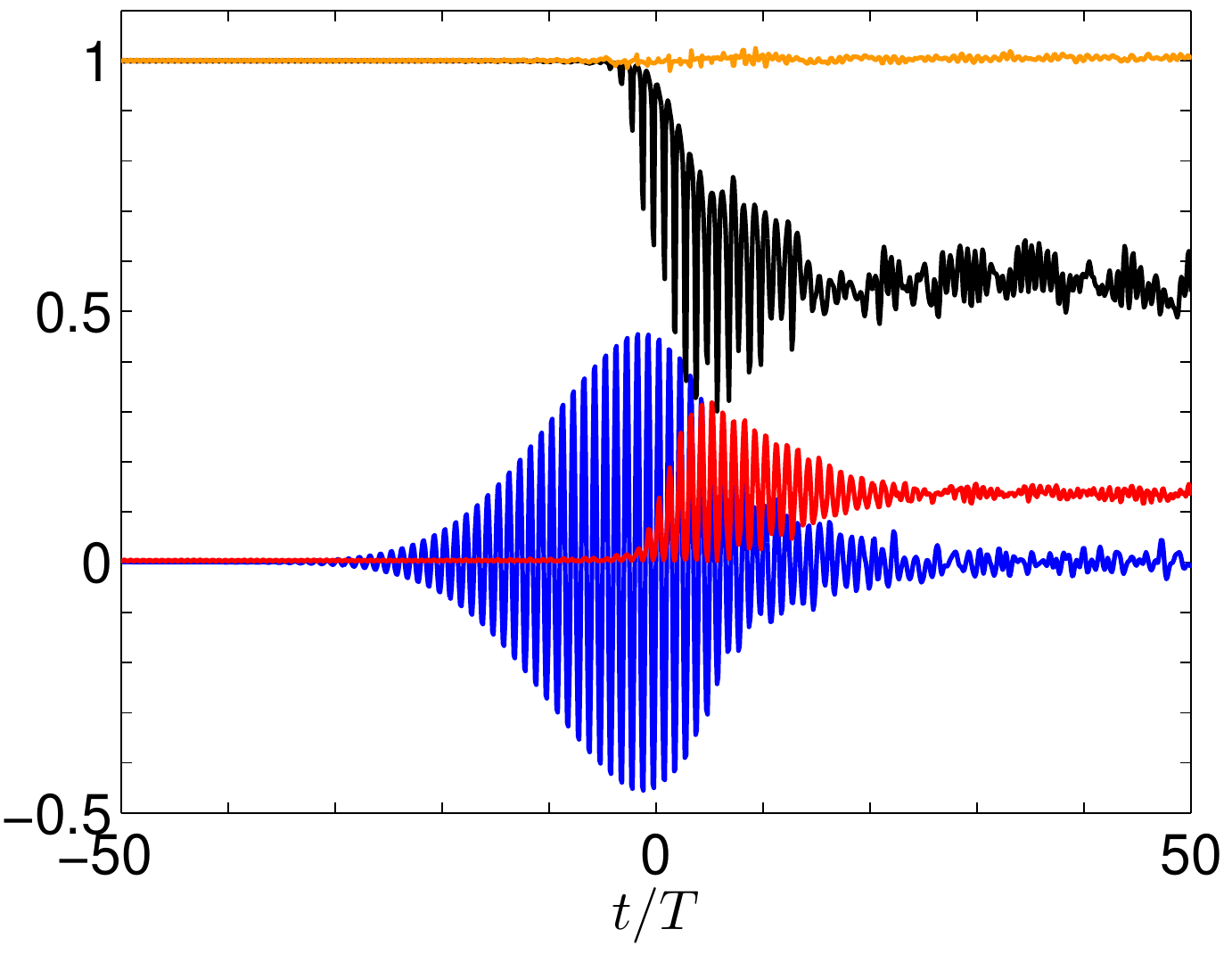}
\end{center}
\caption{(Color online) Dynamics of the driven Josephson 
	junction~(\ref{eq:HDJ}) with $\alpha = 0.95$ and $N = 100$ under the 
	influence of a Gaussian pulse~(\ref{eq:GAP}) with maximum amplitude 
	$\mu_{\rm{max}}/\Omega = 0.9$, scaled frequency $\omega/\Omega = 1.0$, 
	and width $\sigma/T = 10.0$. Shown are 
	the scaled population imbalance $\langle J_z \rangle/N$ (blue line), 
	the closure error $|\Delta_1|/N^{3/2}$ (red line),
	the real part ${\rm Re}(f_1)$ of the ratio~(\ref{eq:RAT}) (orange 
	line), and the stiffness $|R_1|$ (black line). 
	The initial state is the ground state of the undriven 
	junction~(\ref{eq:LMG}).}
\label{F_5}
\end{figure}

\begin{figure}[t]
\begin{center}
\includegraphics[width = 0.9\linewidth]{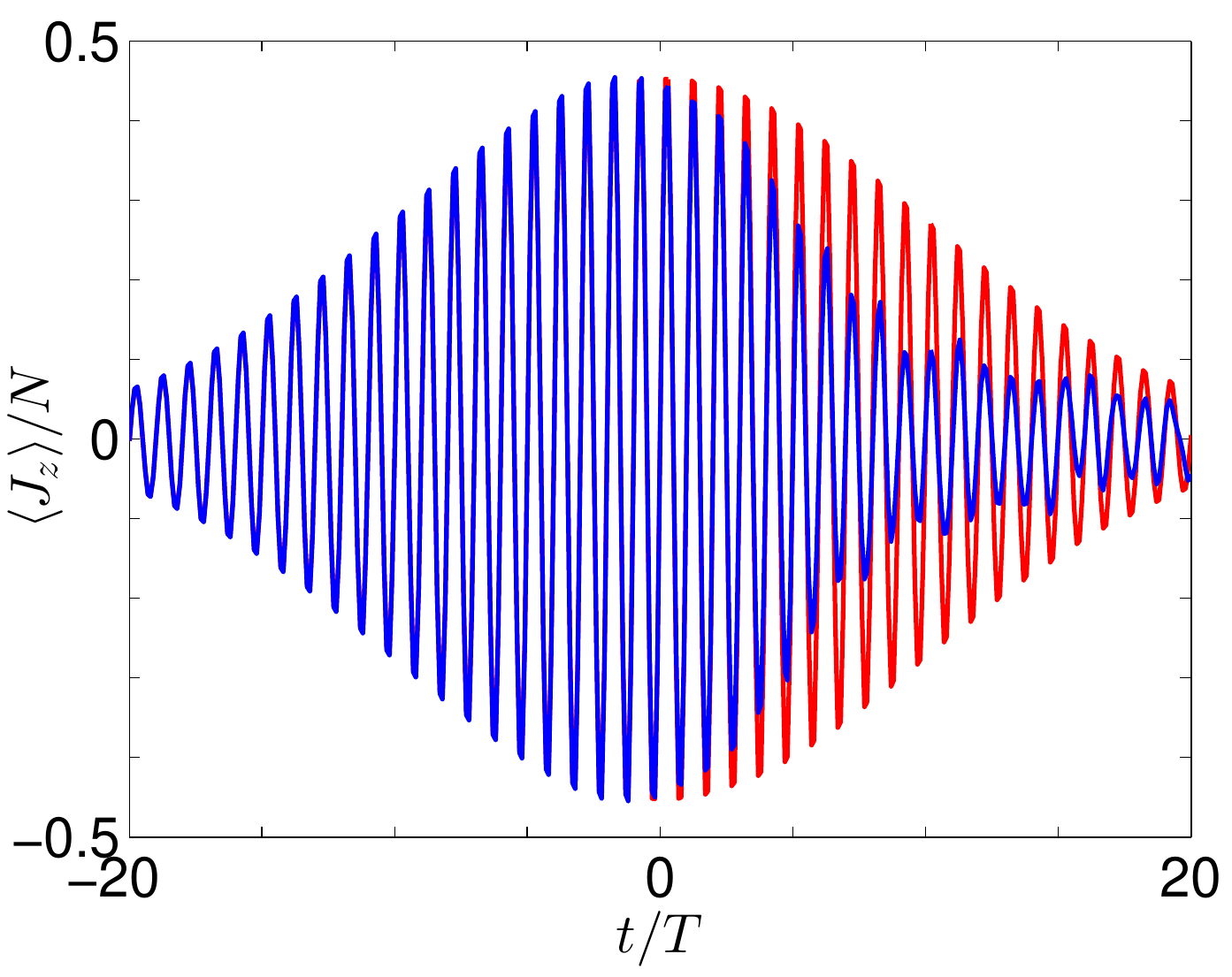}
\end{center}
\caption{(Color online) Comparison of the population imbalance for $N = 100$
	under the action of the pulse monitored in Fig.~\ref{F_5} (blue line)
	to the corresponding prediction of the Gross-Pitaevskii
	equation~(\ref{eq:TSY}) (red line).}   
\label{F_6}
\end{figure}

As initial condition for large negative times we now select the ground state
of the undriven junction~(\ref{eq:LMG}), which, according to Sec.~\ref{sub:31},
differs from an $N$-coherent state even for arbitrarily large~$N$. Hence, now
$|\wP_2(t)\rangle$ is different from $|\wP_1(t)\rangle$, and we have to 
keep track of the ratios~(\ref{eq:RAT}). Figure~\ref{F_5} shows the dynamics
for $\alpha = 0.95$ and comparatively small particle number $N = 100$ in
response to a pulse with maximum scaled amplitude $\mu_{\rm max}/\Omega = 0.9$.
Here the error of closure remains fairly small until a few cycles before the 
pulse's envelope reaches its maximum. During that same interval one finds 
both $f_1(t) \approx 1$ and $f_2(t) \approx 1$ to good accuracy, and, most 
importantly, both stiffnesses $|R_1(t)|$ and $|R_2(t)|$ remain close to
unity before they drop to values fluctuating around about $0.5$ in the 
second half of the pulse. This diagnosis forecasts that the Gross-Pitaevskii 
equation will describe the pulsed $N$-particle dynamics quite faithfully until 
about the pulse's maximum, and then become unreliable. This prediction is 
fully borne out by Fig.~\ref{F_6}, which compares the true population 
imbalance~(\ref{eq:INP}) for this pulse to the mean-field 
imbalance~(\ref{eq:IMF}), as computed from Eq.~(\ref{eq:TSY}): As long as 
$|R_j(t)| \approx 1$, the simple Gross-Pitaevskii equation does a remarkably 
good job.

\begin{figure}[b]
\begin{center}
\includegraphics[width = 0.9\linewidth]{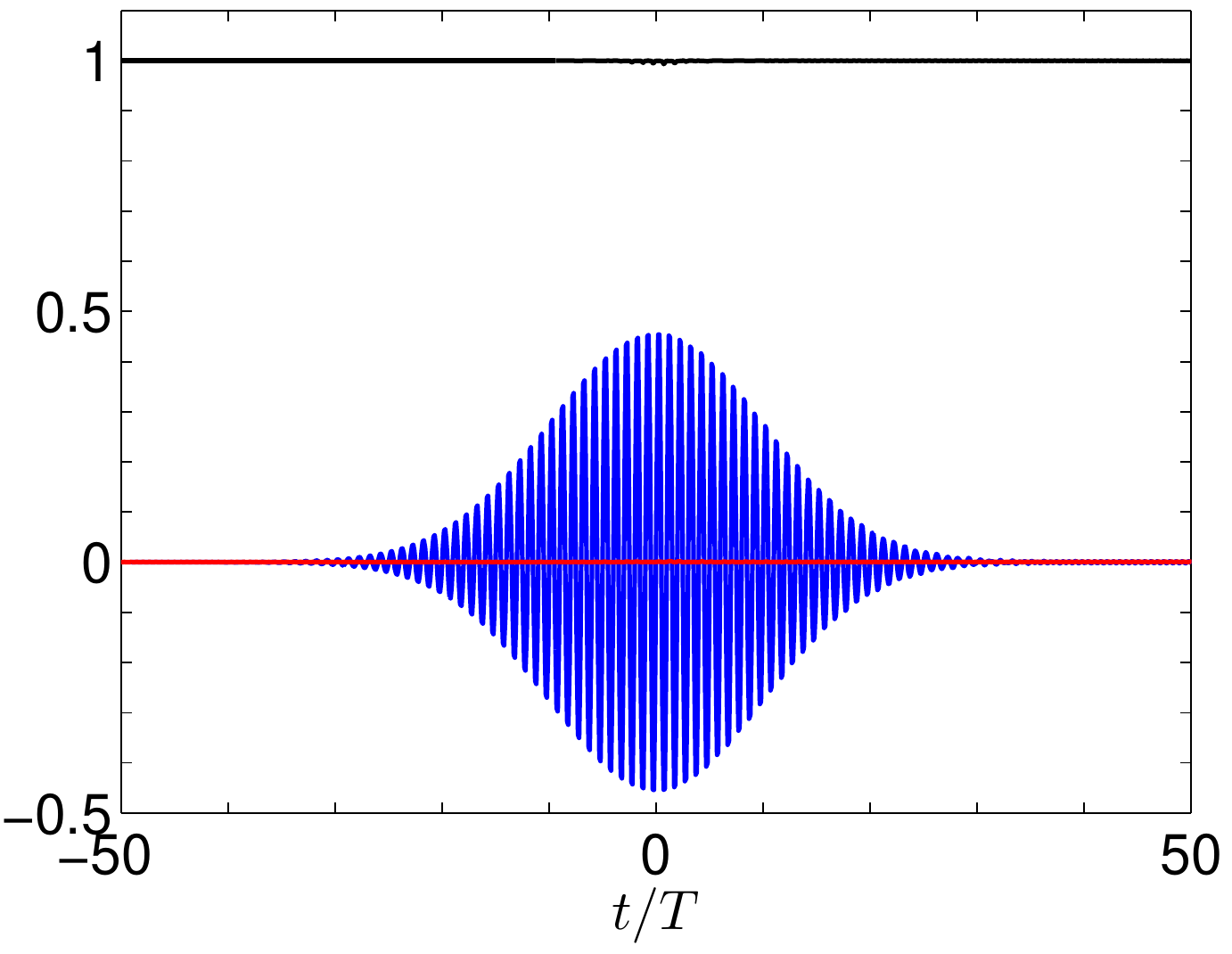}
\end{center}
\caption{(Color online) As Fig.~\ref{F_5}, but with $N = 1000$.}
\label{F_7}
\end{figure}

\begin{figure}[t]
\begin{center}
\includegraphics[width = 0.9\linewidth]{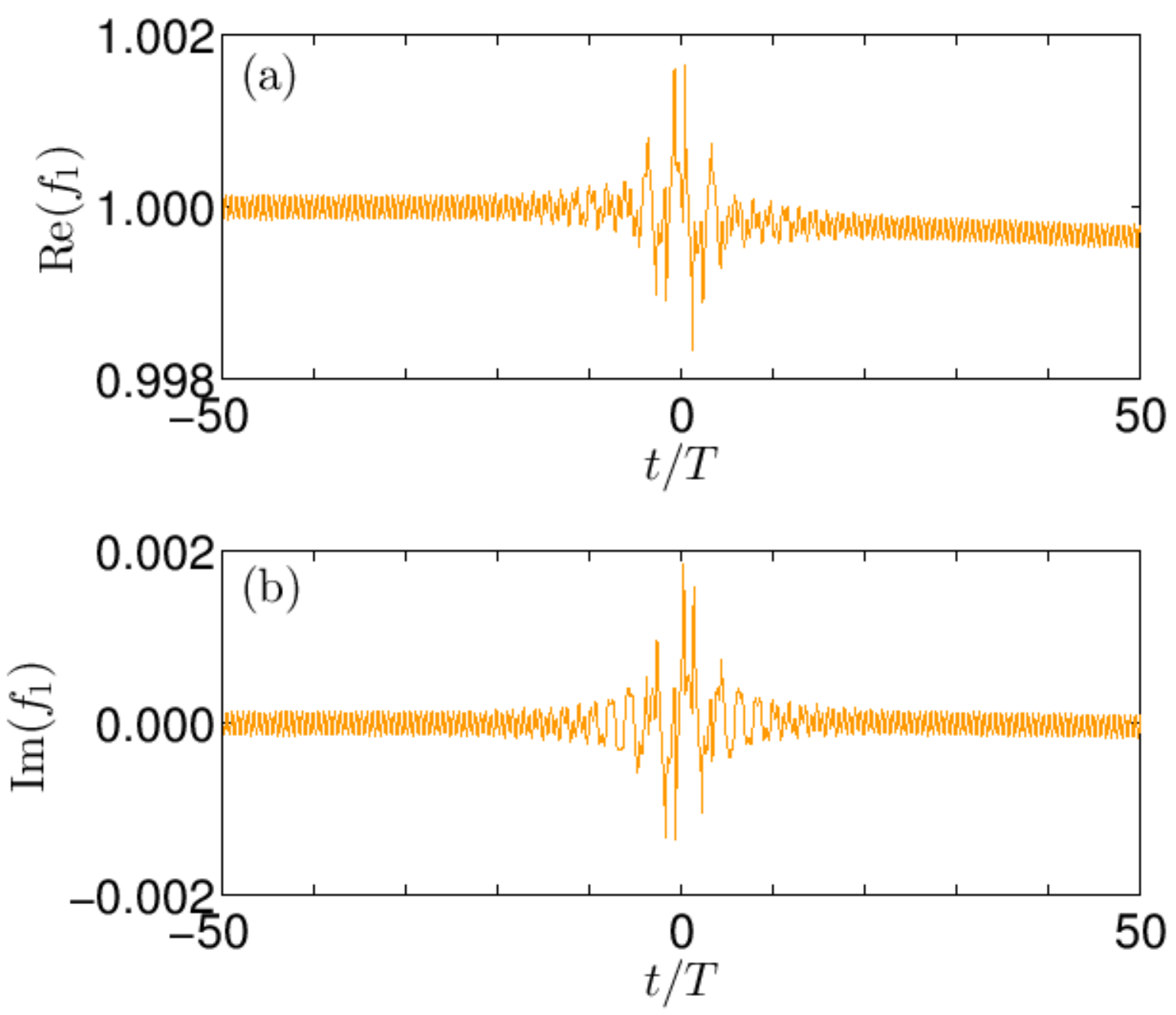}
\end{center}
\caption{(Color online) Real and imaginary part of the ratio $f_1$ 
	for the pulse monitored in Fig.~\ref{F_7}.}
\label{F_8}
\end{figure}

A far more impressive example for the possible accuracy of the 
Gross-Pitaevskii equation is given in Fig.~\ref{F_7}: Here the particle 
number has been increased from $N = 100$ to $N = 1000$, while all other 
parameters are the same as in Fig.~\ref{F_5}, again implying a reduction 
of the interaction strength $\hbar\kappa$ so as to keep~$\alpha$ constant. 
Throughout this pulse the closure error remains negligible, and both ratios 
$f_j(t)$ stay close to unity, as witnessed on a fine scale by Fig.~\ref{F_8}.
This results in almost perfect $t$-coherence of the evolution, even though 
the initial state is not fully $N$-coherent, guaranteeing excellent mean-field 
approximability of the entire pulse dynamics. Since the maximum scaled 
amplitude $\mu_{\rm max}/\Omega = 0.9$ is by no means small, as is evident 
from the quite significant response of the population imbalance, this finding 
is far from trivial.

\begin{figure}[b]
\begin{center}
\includegraphics[width = 0.9\linewidth]{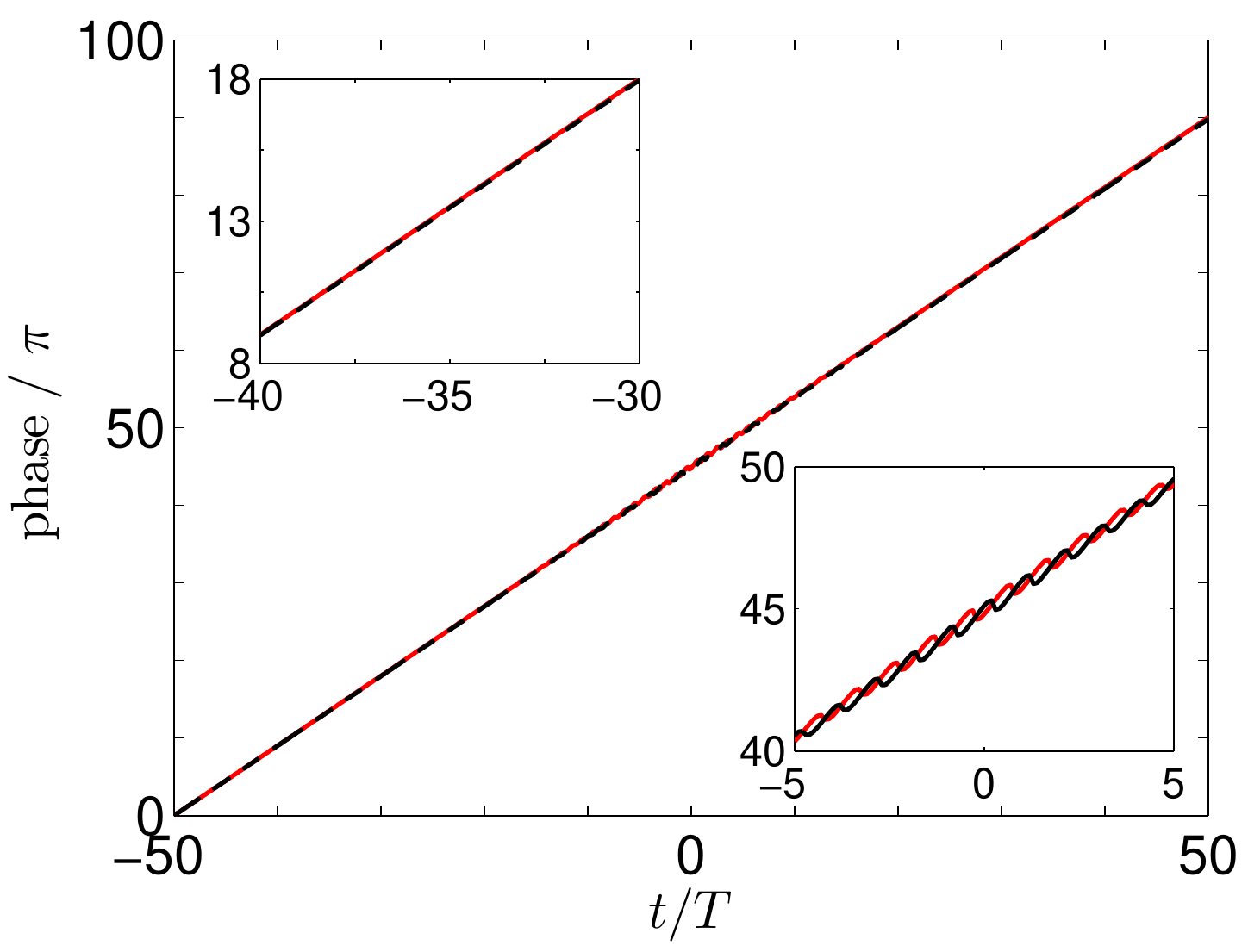}
\end{center}
\caption{(Color online) Comparison of the phase of the mean-field 
	amplitude $c_1(t)$ (red line) with that of $R_1(t)$ (black) for the 
	pulse monitored in Fig.~\ref{F_7}, for which $|R_1(t)| = 1$ to good 
	accuracy. The insets magnify the evolution of the phases at the
	beginning of the pulse, and around its maximum, where in view of 
	Fig.~\ref{F_8} the mean-field approximation is most critical. 
	At the end of the pulse, the phases differ by $0.210 \, \pi$.}
\label{F_9}
\end{figure}

According to Eq.~(\ref{eq:PHD}) the phase of the scalar product $R_j(t)$ 
should determine the phase of the mean-field amplitude $c_j(t)$ in case 
of perfect $t$-coherence, when $| R_j(t) | = 1$. This is confirmed in 
Fig.~\ref{F_9} for the pulse studied in Fig.~\ref{F_7}, which meets the 
above requirement to good accuracy: With $\gamma_1 = 0$, the phase of $c_1(t)$ 
practically coincides with that of $R_1(t)$, growing linearly with time.

\begin{figure}[t]
\begin{center}
\includegraphics[width = 0.9\linewidth]{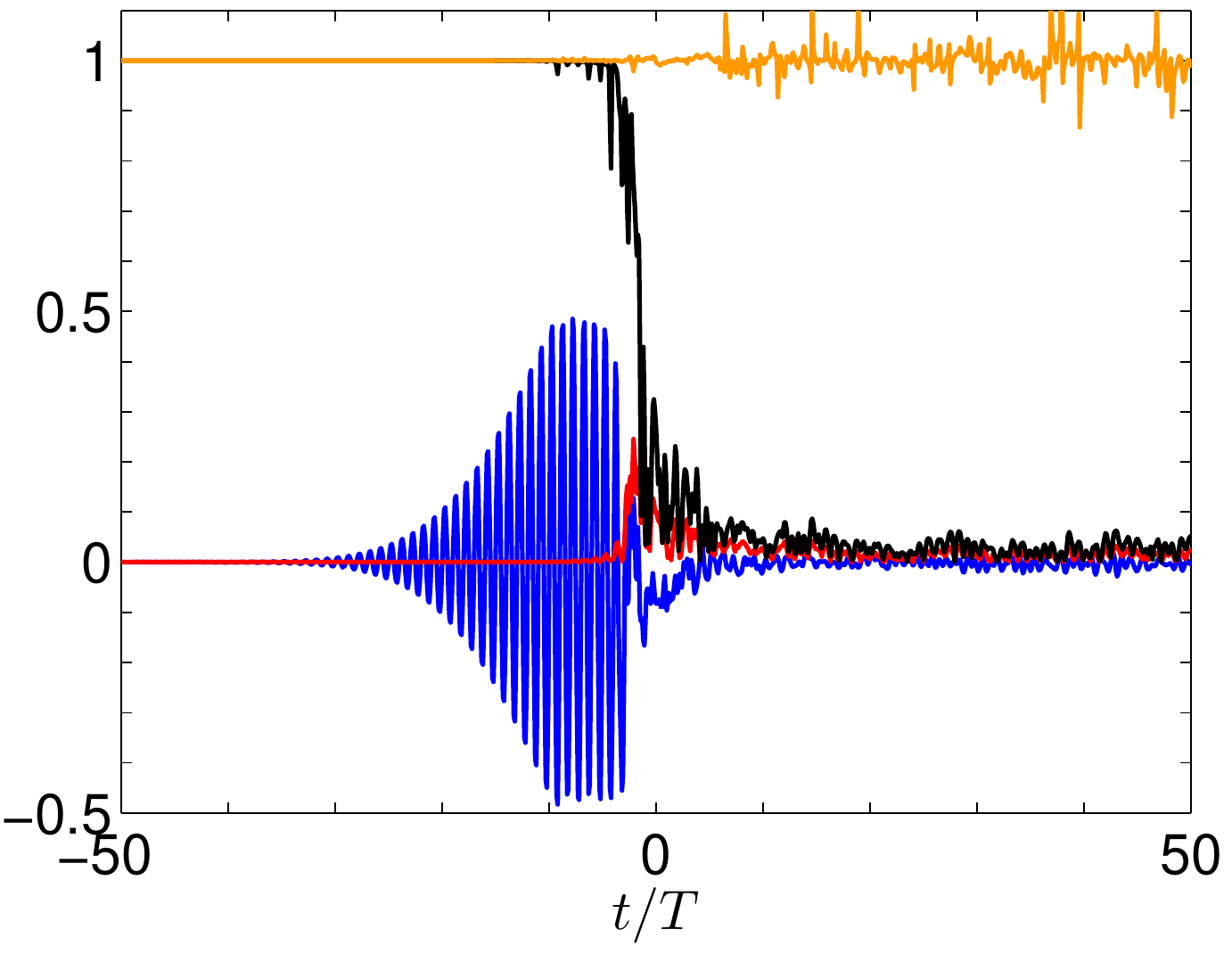}
\end{center}
\caption{(Color online) As Fig.~\ref{F_5}, but with $N = 1000$ and
	$\mu_{\rm{max}}/\Omega = 1.5$.}
\label{F_10}
\end{figure}

\begin{figure}[b]
\begin{center}
\includegraphics[width = 0.9\linewidth]{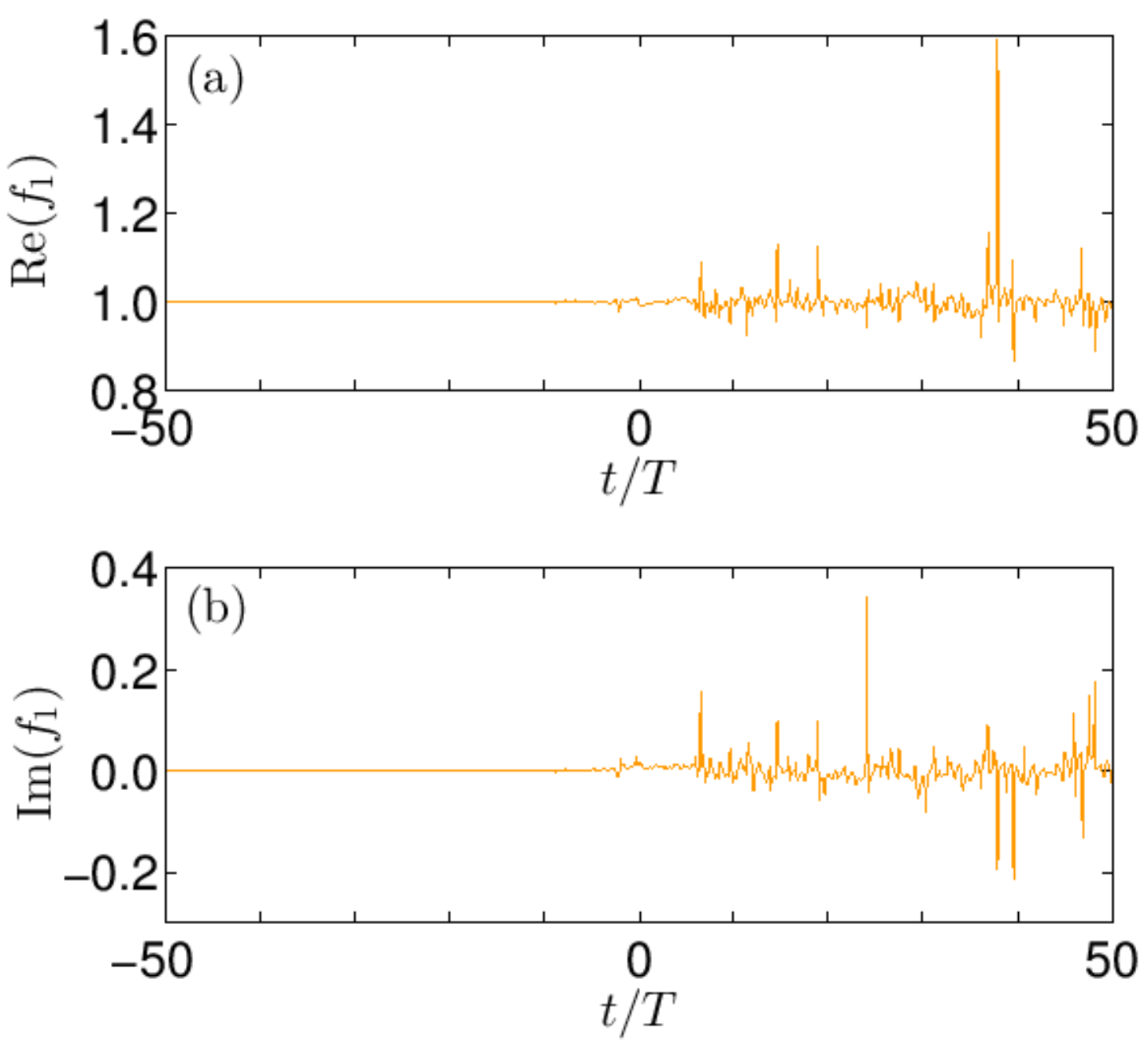}
\end{center}
\caption{(Color online) Real and imaginary part of the ratio $f_1$ 
	for the pulse monitored in Fig.~\ref{F_10}. Observe the change 
	of the ordinates' scale in comparison with Fig.~\ref{F_8}!}
\label{F_11}
\end{figure}

An opposite paradigm is captured by Fig.~\ref{F_10}, again for $N = 1000$, 
but now the maximum scaled amplitude has been increased to 
$\mu_{\rm max}/\Omega = 1.5$. Here the response of the $N$-particle system
is characterized by a stiffness which remains close to perfect almost up 
to the pulse's maximum, but then rapidly drops to values close to zero, as 
a consequence of a suddenly emerging error of closure, and of the sudden 
change of behavior exhibited by the ratios $f_j(t)$ which is detailed in 
Fig.~\ref{F_11}. These errors drive the solution to the Gross-Pitaevskii 
equation~(\ref{eq:TSY}) off its intended track, to such an extent that any 
tangible connection to the actual $N$-particle dynamics appears to be lost.

\begin{figure}[t]
\begin{center}
\includegraphics[width = 0.9\linewidth]{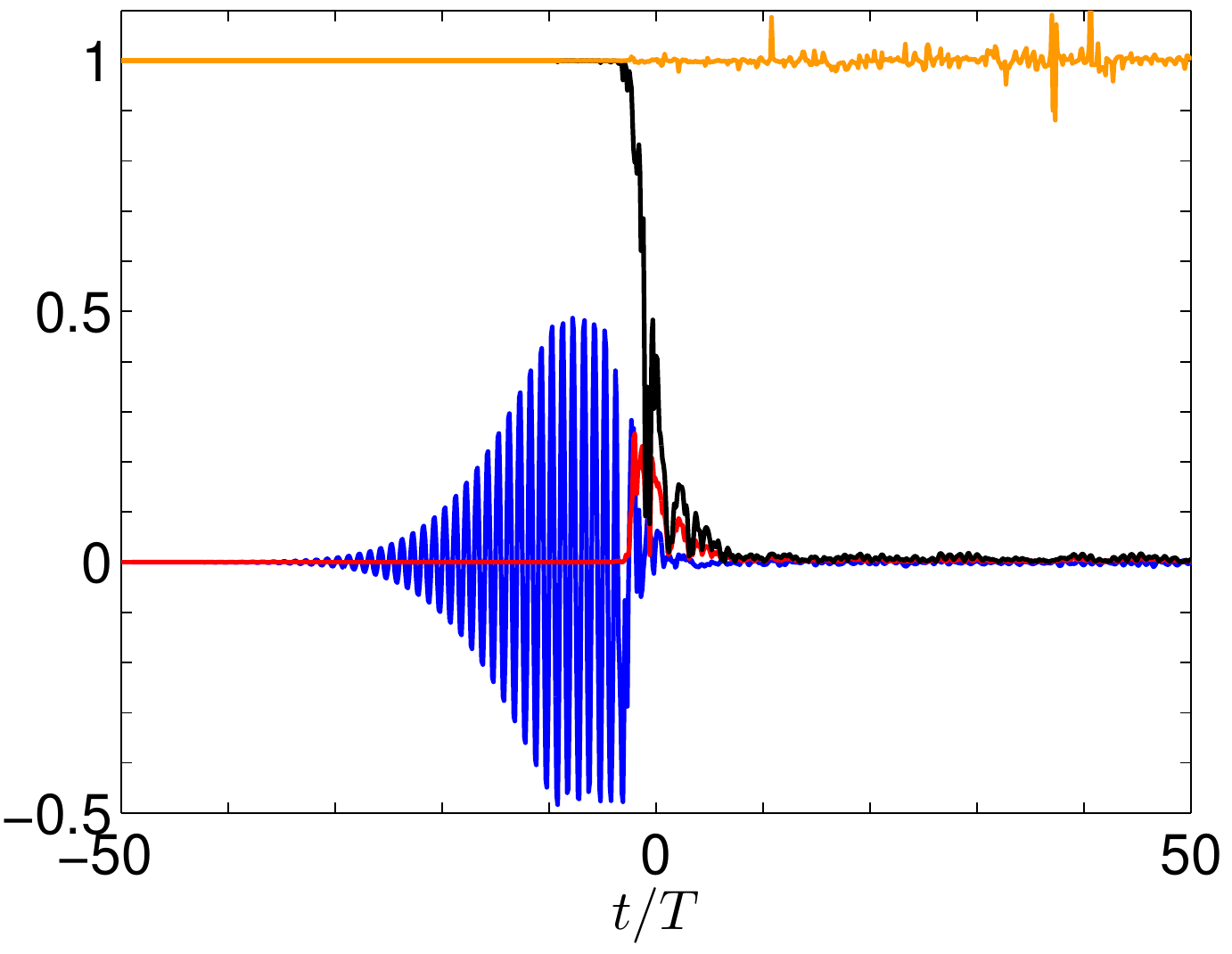}
\end{center}
\caption{(Color online) As Fig.~\ref{F_5}, but with $N = 10000$ and
	$\mu_{\rm{max}}/\Omega = 1.5$, confirming the ``sudden death'' 
	of the mean field already observed in Fig.~\ref{F_10}.}
\label{F_12}
\end{figure}

\begin{figure}[b]
\begin{center}
\includegraphics[width = 0.9\linewidth]{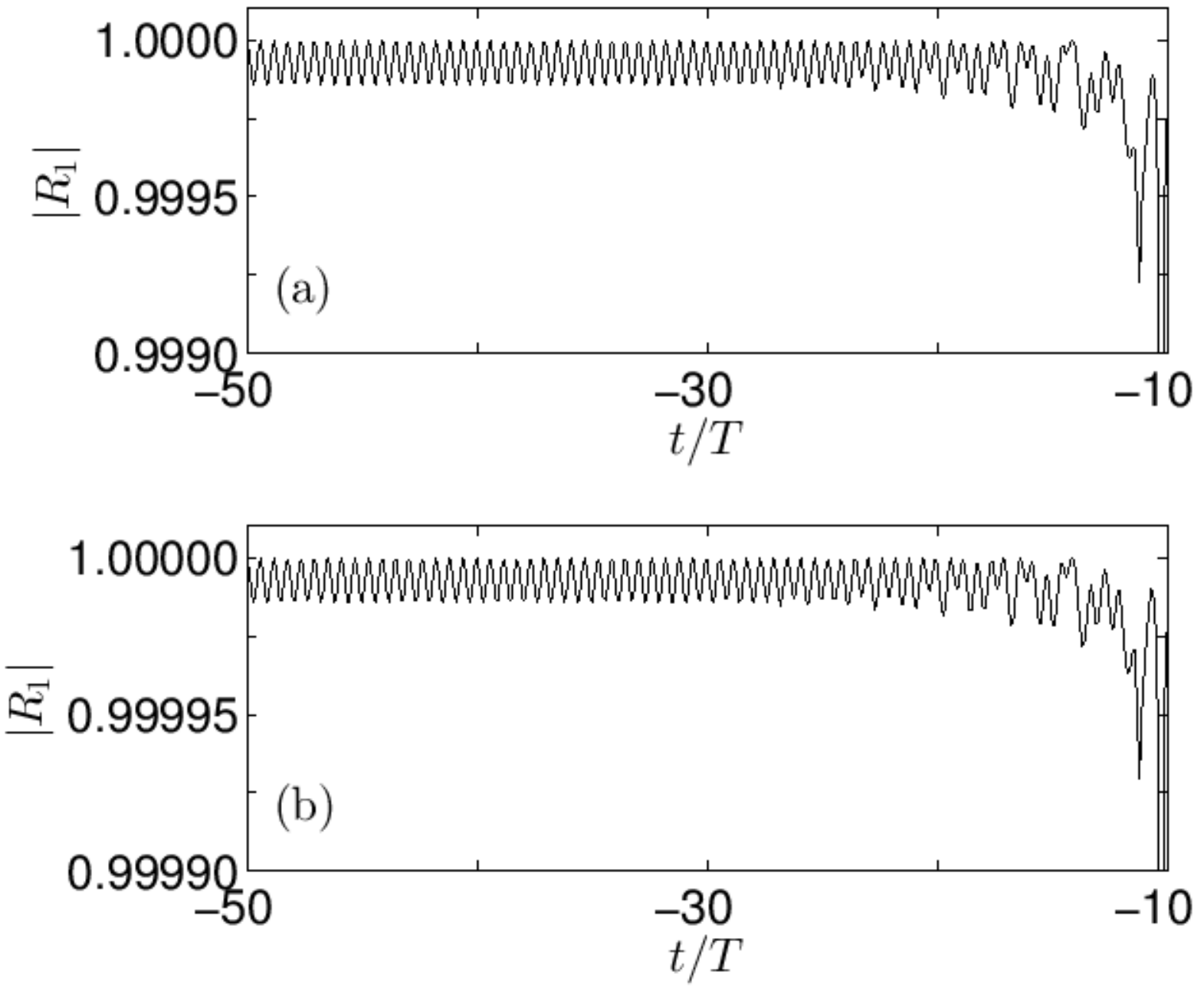}
\end{center}
\caption{Stiffness $|R_1(t)|$ during the rise of the pulses monitored in 
	Fig.~\ref{F_10}, where $N = 1000$ (a), and in Fig.~\ref{F_12}, 
	where $N = 10000$ (b). Observe the scaling with the particle number.}
\label{F_13}
\end{figure}

One might hope that, similar to the effect achieved when going from $N = 100$
in Fig.~\ref{F_5} to $N = 1000$ in Fig.~\ref{F_7}, the loss of $t$-coherence 
recorded in Fig.~\ref{F_10} could be counteracted by increasing the particle
number still further. However, this is {\em not\/} the case: Figure~\ref{F_12}
illustrates the system's response when $N = 10000$, with all other parameters 
equaling those employed in Fig.~\ref{F_10}. While Fig.~\ref{F_13} testifies
that now the stiffness deviates from unity by not more than $2 \times 10^{-5}$ 
during the rise of the pulse, apparently scaling with $1/N$, it then drops even
more sharply, indicating a ``sudden death'' of the mean field, {\em i.e.\/}, a 
dynamically induced sudden loss of $t$-coherence, as corresponding to a sudden 
destruction of the macroscopic wave function~\cite{GertjerenkenHolthaus15}. 
With the help of the following semiclassical deliberations we will argue that 
this loss of $t$-coherence persists in the formal limit $N \to \infty$, when 
$\alpha$ is kept constant: In the situation scrutinized in Figs.~\ref{F_10} 
and~\ref{F_12} there is no chance to describe the system's time evolution 
correctly by means of the Gross-Pitaevskii equation, not even for arbitrarily 
large particle numbers.

\subsection{Semiclassical interpretation}
\label{sub:34}

We now exploit the circumstance that the dynamics generated by the 
time-dependent Gross-Pitaevskii equation~(\ref{eq:TSY}) are equivalent to 
those of a driven nonlinear pendulum~\cite{RaghavanEtAl99,HolthausStenholm01}. 
Starting from the polar representation
\begin{equation}
	c_j = | c_j | \re^{\ri\vartheta_j}
\end{equation}
of the mean-field amplitudes, and introducing their imbalance
\begin{equation}
	z = | c_1 |^2 - | c_2 |^2
\label{eq:DFZ}
\end{equation}
and the relative phase
\begin{equation}
	\phi = \vartheta_2 - \vartheta_1 ,
\end{equation}
the equation of motion~(\ref{eq:TSY}) readily yields
\begin{eqnarray}
	\ri \frac{\rd z}{\rd \tau} & = & c_1 c_2^* - c_1^* c_2
\nonumber \\	& = &
	- | c_1 | | c_2 | \left( \re^{\ri\phi} - \re^{-\ri\phi} \right) \; .
\end{eqnarray}		
Observing
\begin{equation}
	1 - z^2 = 4 | c_1 |^2 | c_2 |^2 \; ,
\end{equation}
this becomes
\begin{equation}
	\frac{\rd z}{\rd \tau} = - \sqrt{1 - z^2} \sin \phi \; .
\label{eq:PEP}
\end{equation}
Similarly, one has
\begin{equation}
	c_1 c_2^* + c_1^* c_2 = \sqrt{1 - z^2} \cos \phi \; ,
\end{equation}
giving		
\begin{equation}
	\frac{\rd}{\rd\tau} \left( c_1 c_2^* + c_1^* c_2 \right)
	= z \sin\phi \cos \phi 
	- \sqrt{1 - z^2} \sin \phi \frac{\rd\phi}{\rd\tau} \; ,
\label{eq:CE1}
\end{equation}		
where Eq.~(\ref{eq:PEP}) has been used. On the other hand, one deduces 
\begin{eqnarray}
	& & 		
	\ri\frac{\rd}{\rd\tau} \left( c_1 c_2^* + c_1^* c_2 \right)
\nonumber \\	& = & 
	2\Big[\alpha \left( | c_1 |^2 - | c_2 |^2 \right) + f(\tau) \Big]
	  \left( c_1 c_2^* - c_1^* c_2 \right)
\end{eqnarray}
from the Gross-Pitaevskii equation~(\ref{eq:TSY}), having written
\begin{equation} 
	f(\tau) = \frac{\mu(\tau)}{\Omega} 
	\sin\left(\frac{\omega}{\Omega} \tau \right) \; ,
\end{equation}
so that
\begin{equation}
	\frac{\rd}{\rd\tau} \left( c_1 c_2^* + c_1^* c_2 \right)
	= -2\Big[ \alpha z + f(\tau) \Big] \sqrt{1 - z^2} \sin\phi \; .
\label{eq:CE2}
\end{equation}	
Combining Eqs.~(\ref{eq:CE1}) and (\ref{eq:CE2}) then leads to 	
\begin{equation}
	\frac{\rd \phi}{\rd \tau} 
	= 2\alpha z + \frac{z}{\sqrt{1 - z^2}} \cos\phi + 2 f(\tau) \; .
\label{eq:PEZ}
\end{equation}		
These equations of motion~(\ref{eq:PEP}) and (\ref{eq:PEZ}) for the 
mean-field imbalance and the relative phase constitute a pair of Hamiltonian 
equations derived from the classical Hamiltonian function
\begin{equation}
	H_{\rm nlp}(z,\phi,\tau) = \alpha z^2 - \sqrt{1 - z^2}\cos \phi 
	+ 2 z f(\tau)
\label{eq:HCL}	 
\end{equation}			
in which $z$ plays the role of a momentum variable, and $\phi$ that of its
canonically conjugate coordinate, and which can thus be interpreted as
the Hamiltonian of a nonlinear pendulum with mass propotional to $1/\alpha$,
and with momentum-dependent length, which is driven by the external force 
$f(\tau)$: Evidently, one has
\begin{equation}
	\frac{\rd \phi}{\rd \tau} = \frac{\partial H_{\rm nlp}}{\partial z}
	\quad , \quad
	\frac{\rd z}{\rd \tau} = -\frac{\partial H_{\rm nlp}}{\partial \phi}
	\; .	
\label{eq:HSY}
\end{equation}
In passing, we remark that for $|z| \ll 1$ this pair reduces to
\begin{eqnarray}
	\frac{\rd \phi}{\rd \tau} & \approx & 2 f(\tau)
\nonumber \\
	\frac{\rd z}{\rd \tau} & \approx & -\sin\phi \; ,
\end{eqnarray}
as corresponding to the equations for the phase and the current across
a macroscopic superconducting Josephson junction~\cite{BaronePaterno82}.
 
This classical viewpoint underlines the significance of the extension
of the Lipkin-Meshkov-Glick model~(\ref{eq:LMG}) by the time-dependent
drive~(\ref{eq:HDR}): When $f(\tau) = 0$, the Hamiltonian~(\ref{eq:HCL})
represents a dynamical system with one single degree of freedom, possessing 
energy as its integral of motion. Adding a time-dependent force is tantamount 
to adding a further degree of freedom not accompanied, in general, by a further
integral of motion, so that the driven nonlinear pendulum is non-integrable in 
the sense of classical Hamiltonian mechanics~\cite{Gutzwiller90,JoseSaletan98},
giving rise to chaotic motion. This raises the question how the actual, linear 
$N$-particle system behaves when its nonlinear mean-field descendant becomes 
chaotic.

\begin{figure*}[t]
\begin{center}
\includegraphics[width = 0.9\linewidth]{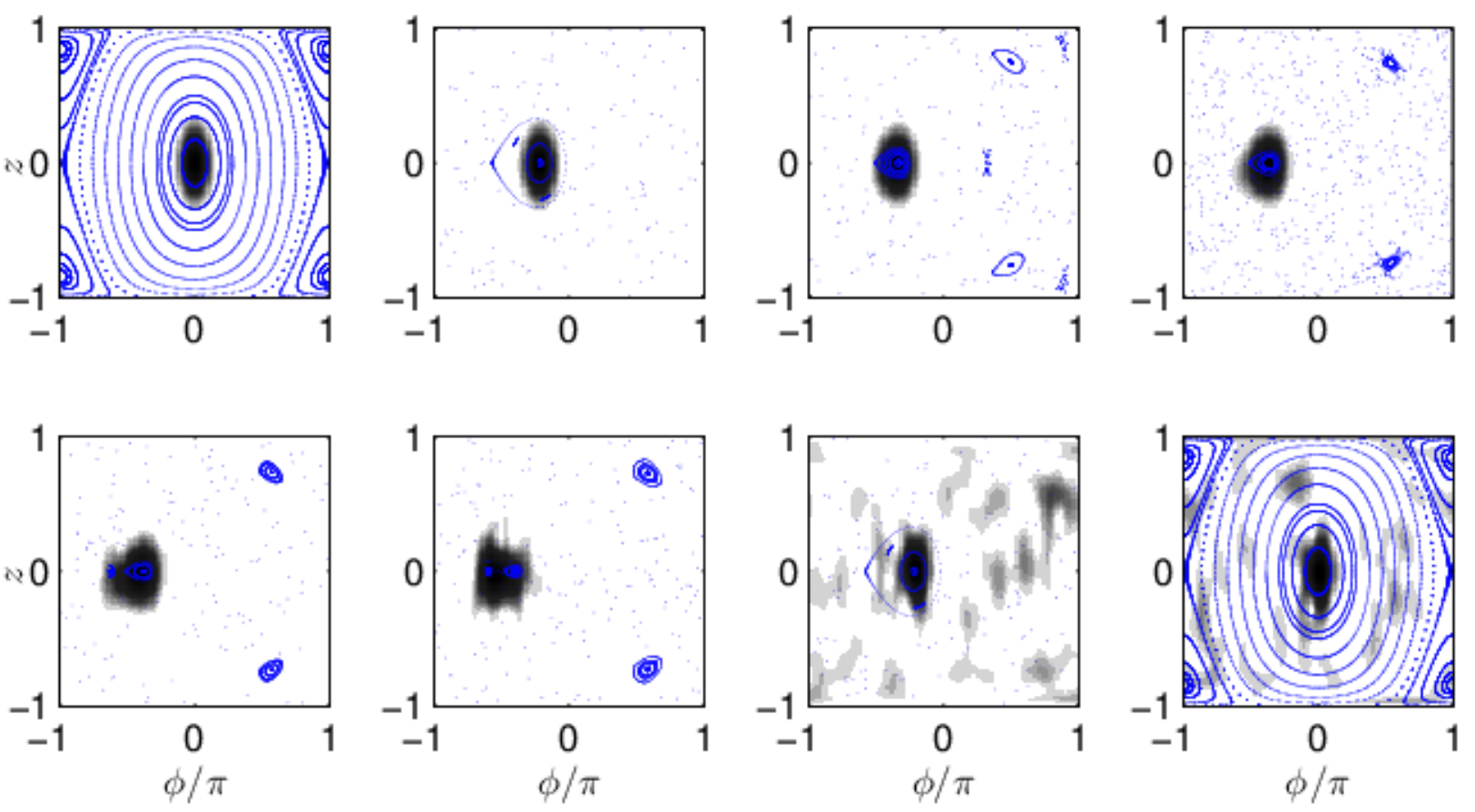}
\end{center}
\caption{(Color online) Grey-scale Husimi snapshots of the evolving 
	$N$-particle state for the pulse monitored in Fig.~\ref{F_5}, for 
	which $N = 100$ and $\mu_{\rm max}/\Omega = 0.9$, superimposed onto 
	the instantaneous Poincar\'e surfaces of section (blue) at successive 
	times~$t$ when $\mu(t)/\Omega = 0.0$, $0.6$, $0.83$, $0.86$, $0.88$, 
	$0.9$, $0.6$, $0.0$ (upper left to lower right).
	Here the quantum state initially remains associated with an
	adiabatic invariant corresponding to a closed curve in the regular
	island, as selected by the Bohr-Sommerfeld condition~(\ref{eq:BSC})
	with $n = 0$, but then spreads over the stochastic sea when the 
	island becomes so small that the invariant is destroyed.} 
\label{F_14}
\end{figure*}

A useful link between the $N$-particle level and its mean-field description 
in terms of the Hamiltonian system~(\ref{eq:HSY}) is provided by the
$SU(2)$-coherent states~(\ref{eq:ACS}): Taking the expectation value of
the $N$-particle Hamiltonian~(\ref{eq:HDJ}) with respect to these states,
one obtains 
\begin{eqnarray} & & 
	\frac{{_N\langle} \theta,\phi | H(t) | \theta,\phi \rangle_N}
	     {N\hbar\Omega/2}
\nonumber \\ & = & 
	\alpha\left( 1 - \frac{1}{N}\right) \left(\cos^2\theta + 1 \right) 	
	-\sqrt{1 - \cos^2\theta} \cos\phi
\nonumber \\ & & 	
	+ 2 \cos\theta \; \frac{\mu(t)}{\Omega}\sin(\omega t) \; , 		
\end{eqnarray}
which equals the classical Hamiltonian~(\ref{eq:HCL}) up to an irrelevant
constant if one neglects terms of order ${\mathcal O}(1/N)$, and sets 
\begin{equation}
	z = \cos\theta \; .
\end{equation}	
Since $\cos\theta = \cos^2(\theta/2) - \sin^2(\theta/2)$, this is in 
perfect correspondence with the representation~(\ref{eq:NSP}) of the 
$N$-fold occupied single-particle state which underlies the $N$-coherent 
state $| \theta,\phi \rangle_N$ on the one hand, and with the 
definition~(\ref{eq:DFZ}) on the other. Therefore, the squared projections
\begin{equation}
	\mathcal{Q}_N(z,\phi,t) = 
	\big| {_N\langle} \theta,\phi | \Psi(t) \rangle \big|^2 \; , 
\label{eq:HUD}
\end{equation}
referred to as Husimi distributions, quantify the degree of affinity of the 
$N$-particle state $| \Psi(t) \rangle$ with the phase-space point $(z,\varphi)$
at the moment~$t$~\cite{WeissTeichmann08,GraefeEtAl14,GertjerenkenHolthaus14}.
This observation leads to the desired semiclassical view on the pulse dynamics 
recorded in Sec.~\ref{sub:33}: We take the solution $| \Psi(t) \rangle$ to 
the $N$-particle Schr\"odinger equation at certain moments~$t$, compute the 
associated Husimi distributions~(\ref{eq:HUD}), and superimpose these 
distributions onto the corresponding classical phase-space portraits, that is, 
onto the Poincar\'e surfaces of section of the accompanying pendulum which 
is periodically driven with amplitude $\mu/\Omega$ ``frozen'' at the value 
reached at the moment~$t$ under study. These surfaces of section are computed 
by solving the classical equations of motion for a representative set of 
initial conditions with fixed driving amplitude~$\mu/\Omega$, and by plotting 
the resulting phase-space points stroboscopically after each driving 
period~$2\pi/\omega$.

\begin{figure*}[t]
\begin{center}
\includegraphics[width = 0.9\linewidth]{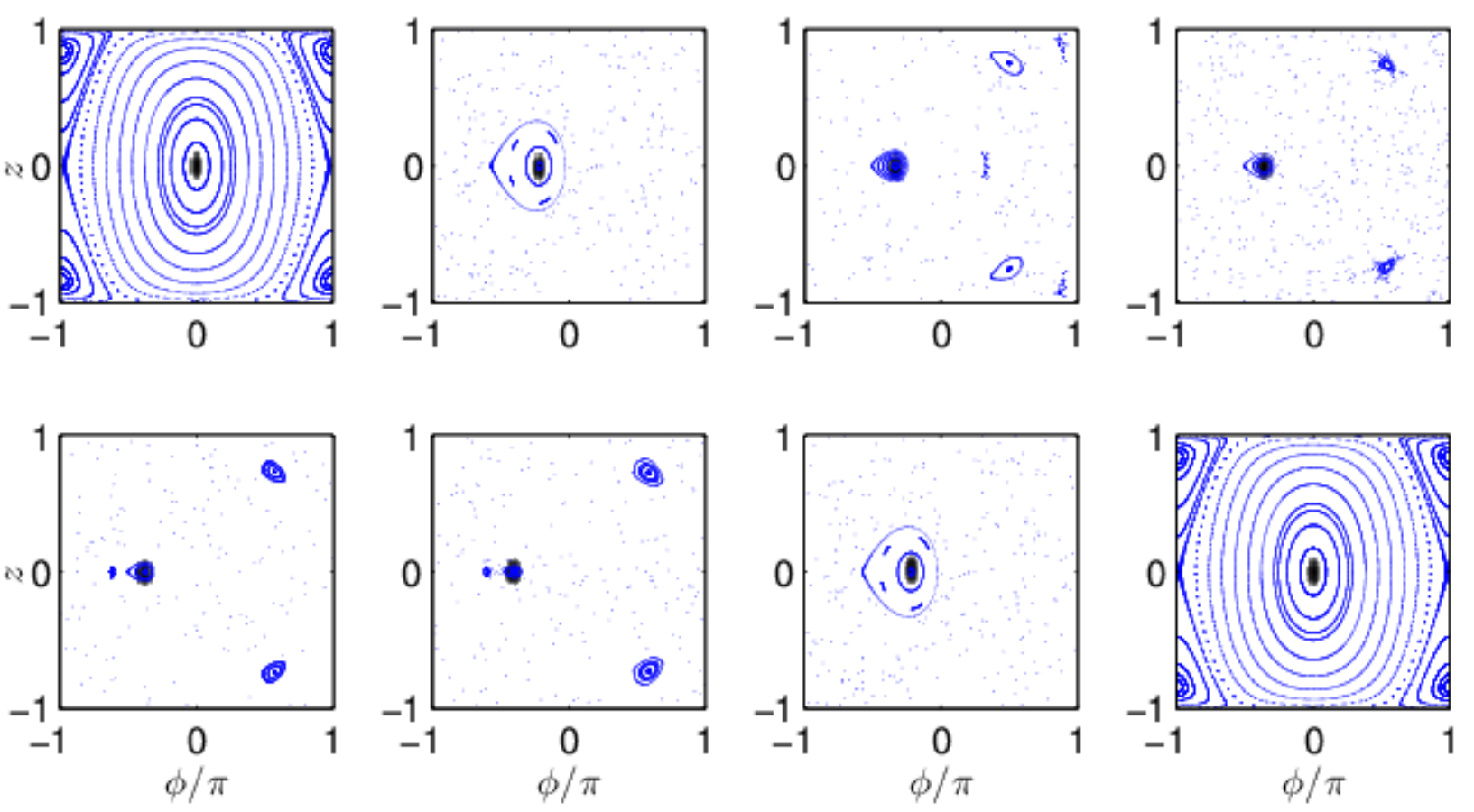}
\end{center}
\caption{(Color online) Grey-scale Husimi snapshots of the evolving 
	$N$-particle state for the pulse monitored in Fig.~\ref{F_7}, for 
	which $N = 1000$ and $\mu_{\rm max}/\Omega = 0.9$, superimposed onto 
	the instantaneous Poincar\'e surfaces of section (blue) at successive 
	times~$t$ when $\mu(t)/\Omega = 0.0$, $0.6$, $0.83$, $0.86$, $0.88$, 
	$0.9$, $0.6$, $0.0$ (upper left to lower right). While the Poincar\'e 
	sections are the same as those displayed in Fig.~\ref{F_14}, 
	now the curve~$\gamma_0$ required by the Bohr-Sommerfeld 
	condition~(\ref{eq:BSC}) encircles an area which is reduced by a 
	factor of $10$, and thus remains confined in the regular island 
	during the entire pulse, enabling adiabatic following of the 
	$N$-particle quantum state.}
\label{F_15}
\end{figure*}

Figure~\ref{F_14} shows a series of snapshots obtained in this manner for 
the pulse previously studied in Fig.~\ref{F_5}, for which $N = 100$ and 
$\mu_{\rm max}/\Omega = 0.9$. Initially, when $\mu/\Omega = 0$, the 
accompanying classical pendulum is integrable, as is reflected by a 
phase-space portrait which is stratified into closed curves remaining 
invariant under the Hamiltonian flow. The initial quantum state considered 
here, which is the ground state of the undriven system~(\ref{eq:LMG}) as 
depicted in Fig.~\ref{F_1}, is semiclassically linked by means of the 
WKB-construction to that invariant curve $\gamma_n$ surrounding the 
central fixed point which is selected by the Bohr-Sommerfeld 
condition~\cite{NissenKeeling10,SimonStrunz12,GraefeEtAl14,
GertjerenkenHolthaus14}  
\begin{equation} 
	\frac{1}{2\pi}\oint_{\gamma_n} \! z\mathrm{d}\phi = 
	\hbar_{\rm eff}\left(n + \frac{1}{2}  \right)	
\label{eq:BSC}
\end{equation}
with effective Planck constant
\begin{equation}
	\hbar_{\rm eff} = \frac{2}{N} \; ,	
\label{eq:EPC}
\end{equation} 
and with quantum number $n = 0$, so that its Husimi distribution appears
concentrated around that curve $\gamma_0$. When the driving amplitude is
increased, a large fraction of the invariant curves is destroyed giving way
to chaotic motion, while those surrounding the central fixed point are
deformed, but remain preserved if $\mu/\Omega$ does not become too large, 
forming a regular island embedded in a chaotic sea. If the pulse's amplitude 
increases on a time scale which is slow compared to the period $2\pi/\omega$, 
these preserved curves represent adiabatic invariants to which the 
time-evolving wave function remains tied in a WKB-type manner. This is what 
explains the scenario depicted in Fig.~\ref{F_14}: For low driving strength 
the evolving quantum state clinges to its adiabatic invariant still contained 
in the regular island, but with increasing amplitude the island becomes so 
small that the required invariant does no longer exist. Then the Husimi 
projection of the quantum state, having nothing left it can cling to, spills 
out more or less over the entire phase space, leading to a final quantum state 
which contains many eigenstates of the undriven junction~(\ref{eq:LMG}).
  
With this background knowledge the difference between the pulses recorded in 
Fig.~\ref{F_5} on the one hand, and in Fig.~\ref{F_7} on the other, becomes
intuitively clear, recalling that for the latter pulse the maximum scaled 
amplitude $\mu_{\rm max}/\Omega = 0.9$ has been maintained while the particle 
number has been increased to $N = 1000$. Because this implies that the 
effective Planck constant~(\ref{eq:EPC}) is reduced by a factor of~$10$, the 
required phase-space curve $\gamma_0$ now encircles a correspondingly smaller 
area. As shown in Fig.~\ref{F_15}, it therefore fits into the shrinking and 
re-growing regular island during the entire pulse, albeit just barely so at 
its maximum. Thus, the quantum state evolving under the influence of the 
pulse remains semiclassically associated with an adiabatic invariant which is 
preserved from the pulse's beginning until its end, and thus enables adiabatic 
following. As a result, the final $N$-particle state here closely resembles 
the initial one.

\begin{figure*}[t]
\begin{center}
\includegraphics[width = 0.9\linewidth]{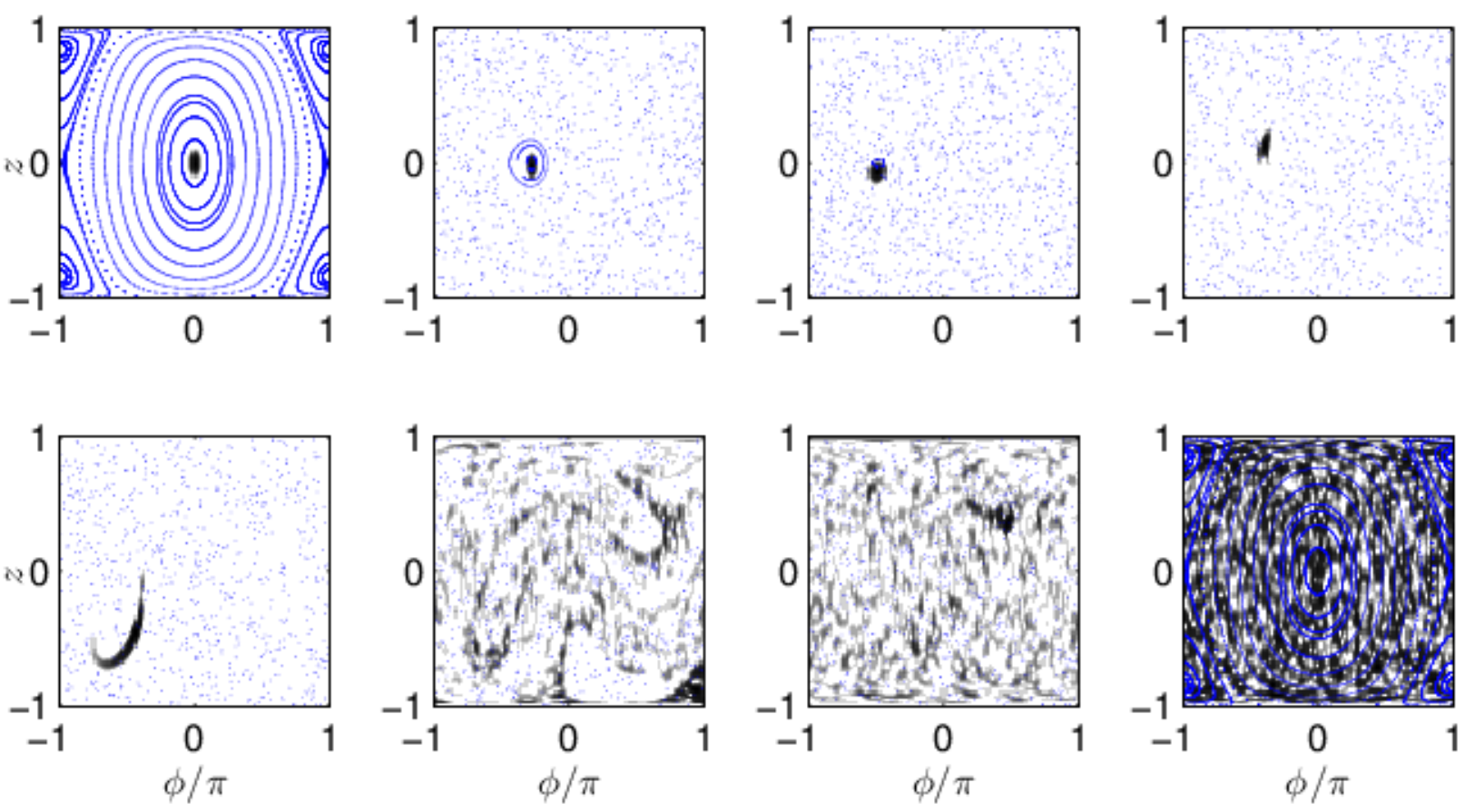}
\end{center}
\caption{(Color online) Grey-scale Husimi snapshots of the evolving 
	$N$-particle state for the pulse monitored in Fig.~\ref{F_10}, for 
	which $N = 1000$ and $\mu_{\rm max}/\Omega = 1.5$, superimposed onto 
	the instantaneous Poincar\'e surfaces of section (blue) at successive 
	times~$t$ when $\mu(t)/\Omega = 0.0$, $0.73$, $1.0$, $1.25$, $1.43$, 
	$1.5$, $1.43$, $0.0$ (upper left to lower right). Here the regular 
	island is completely destroyed in the middle of the pulse, implying 
	that there is no adiabatic invariant capable of carrying a macroscopic 
	wave function. Thus, even for arbitrarily large particle number the 
	state becomes so complex that a Gross-Pitaevskii description is 
	impossible.}       
\label{F_16}
\end{figure*}

A mere glimpse on Fig.~\ref{F_16} then suffices to grasp why no further 
enhancement of the particle number whatsoever, together with the implied 
reduction of the inter\-particle interaction in order to maintain a constant 
value of~$\alpha$, could possibly bring one back to the Gross-Pitaevskii 
regime under the conditions of Figs.~\ref{F_10} and \ref{F_12}: Here the pulse 
possesses the maximum amplitude $\mu_{\rm max}/\Omega = 1.5$, which is so 
strong that the regular island which still hosts the adiabatic invariant at the
early stages of the pulse is destroyed completely when the envelope approaches 
its maximum. Hence, no matter how large the particle number~$N$ and, 
consequently, how small the effective Planck constant~(\ref{eq:EPC}), there is 
no preserved adiabatic invariant which could possibly carry a macroscopic wave 
function. Instead, Fig.~\ref{F_16} visualizes that upon destruction of the 
island the quantum state aquires a degree of complexity which strikingly 
contrasts the simplicity of the initial condition. Obviously this sudden
increase of complexity, as corresponding to a sudden loss of order, reflects 
the sudden drop of stiffness observed in Figs.~\ref{F_10} and \ref{F_12}, 
signaling the sudden death of the mean field. We suggest that this 
scenario is quite general, so that it would be experimentally accessible, 
for instance, with pulsed Bose-Einstein condensates in anharmonic trapping 
potentials~\cite{GertjerenkenHolthaus15}.

\section{Discussion}
\label{sec:4}

One usually takes recourse to a mean-field theory, which in many cases is still
tractable by numerical means, to obtain information on intricate many-particle 
dynamics, which in general do not lend themselves to direct simulations. In our 
approach to the time-dependent Gross-Pitaevskii equation this common strategy 
has been reversed, in order to clarify questions concerning the emergence 
and possible persistence of an order parameter in a system of $N$~Bosons 
exposed to external forcing: Assuming that we know its exact many-particle 
wave function $\Psi(\bm r_1, \ldots, \bm r_N; t)$, we have asked how this 
knowledge can be used for constructing the associated macroscopic wave 
function, provided that it exists, in the form of a mean-field amplitude 
$\Phi\rt$. The answer, formalized in Eqs.~(\ref{eq:AUI}), (\ref{eq:AES}), 
(\ref{eq:AUF}), and (\ref{eq:DEF}), is conceptually interesting insofar as 
it does neither involve the formal limit $N \to \infty$, nor the notion of 
spontaneous symmetry breaking: Take the initial $N$-particle state, annihilate 
a particle, and propagate both the initial state and the subsidiary 
$(N-1)$-particle states thus obtained in time; a candidate mean-field amplitude
then is introduced by taking the matrix elements of the annihilation operator 
with these evolving states. 

The starting point of this construction resembles the specification of a 
condensate wave function by Lifshitz and Pitaevskii~\cite{LaLifIX}, but the 
additional consideration of time evolution in response to external forcing 
brings about a new twist. Namely, in order to decide whether or not the 
candidate actually is a proper mean-field amplitude, which is tantamount to
the question whether or not a macroscopic wave function of the driven system 
still does exist, it is not sufficient to follow solely the trajectory of the 
given $N$-particle state in Fock space. Rather, one has to compare this given
trajectory to other trajectories initially close to it, differing by one in
particle number, and to check to what extent these initially close trajectories
diverge in the course of time. This is what is effectively quantified by the
projection~(\ref{eq:DCO}): If the absolute value of this scalar product~$R\rt$
equals unity to good approximation, the state obtained by first evolving 
the initial condition in time and annihilating a particle at a later moment 
differs merely by a phase factor from the states obtained by annihilating first
and evolving thereafter. This property of $t$-coherence, indicated by 
$|R\rt| = 1$, means that initial states differing by one in particle number 
move in some sense parallel to each other in Fock space. In this case the flow 
in Fock space can be considered as stiff, similar to laminar flow in fluid 
mechanics. This is the quality of ``simplicity'' required for ensuring the 
presence of a genuine mean-field amplitude: If $|R\rt| = 1$ at least to a good 
approximation, the candidate actually qualifies as a macroscopic wave function 
and obeys the time-dependent Gross-Pitaevskii-equation; if not, it has no 
immediate physical significance.

The model of the driven Josephson junction studied in Sec.~\ref{sec:3} 
suggests that the occurrence of chaotic solutions to the Gross-Pitaevskii 
equation reflects the loss of $t$-coherence on the $N$-particle level.
Our findings thus extend previous studies~\cite{CastinDum97,GardinerEtAl00,
BrezinovaEtAl11,BrezinovaEtAl12} which have cast doubt on the validity of 
a Gross-Pitaevskii-type mean-field approach under chaotic conditions. It 
appears that the distinction between order and chaos, which has been explored 
in great depth in classical mechanics~\cite{Gutzwiller90,JoseSaletan98}, may 
have implications of its own in quantum many-body physics. In particular, the 
scenario depicted in Fig.~\ref{F_12} indicates in a striking manner that a 
macroscopic wave function can be destroyed almost instantaneously upon 
entering a chaotic regime.      

Recent pioneering works which have taken up the investigation of many-body
quantum chaos have considered $\delta$-kicked condensates~\cite{ZhangEtAl04,
LiuEtAl06,DuffyEtAl04,WimbergerEtAl05,BillamGardiner12,BillamEtAl13}. In
contrast, here we have studied the response to a sinusoidal force with a 
``slowly'' varying envelope. One of our most important results, albeit 
obtained for one particular model system only, consists in the observation 
that adiabatic following to such slowly changing driving forces allows 
one to preserve a pre-existing macroscopic wave function, and to transport 
it almost without loss of $t$-coherence into the regime of strong driving, 
as shown exemplarily in Fig.~\ref{F_7}. This finding almost provides a 
blueprint for generating Floquet condensates~\cite{GertjerenkenHolthaus14}. 
More generally, it may be of interest for guiding further experiments with 
time-periodically forced Bose-Einstein condensates intended to engineer novel 
systems which may not be accessible without such forcing~\cite{EckardtEtAl05,
EckardtEtAl09,ZenesiniEtAl09,StruckEtAl11,MaEtAl11,ChenEtAl11,HaukeEtAl12,
StruckEtAl13,ParkerEtAl13,AidelsburgerEtAl13,GoldmanDalibard14,Eckardt15}: 
If it is possible to preserve maximum stiffness, or $t$-coherence, even in the 
presence of strong forcing, it should also be possible to actively manipulate 
macroscopic wave functions by applying suitable coherent control techniques. 
The evidence collected in the present work clearly indicates that this road 
is viable.

\begin{acknowledgments}
We thank S.~Arlinghaus, M.~Tschikin, and C.~Weiss for discussions during 
the early stages of this work. We also acknowledge support from the 
Deutsche Forschungs\-gemeinschaft (DFG) through grant No.\ HO 1771/6-2. 
The computations were performed on the HPC cluster HERO, located at the 
University of Oldenburg and funded by the DFG through its Major Research 
Instrumentation Programme (INST 184/108-1 FUGG), and by the Ministry of 
Science and Culture (MWK) of the Lower Saxony State.
\end{acknowledgments}


\begin{thebibliography}{99}

\bibitem{HuangYang57} 
	K. Huang and C. N. Yang,
	Phys. Rev. {\bf 105}, 767 (1957).

\bibitem{Gross61} 
	E. P. Gross,
	Nuovo Cimento {\bf 20}, 454 (1961).
	
\bibitem{Pitaevskii61} 
	L. P. Pitaevskii,
	Zh. Eksp. Teor. Fiz. {\bf 40}, 646
	[Sov. Phys. JETP {\bf 13}, 451] (1961).	

\bibitem{Gross63} 
	E. P. Gross,
	J. Math. Phys. {\bf 4}, 195 (1963).
	
\bibitem{Gardiner97}
	C. W. Gardiner,
	Phys. Rev. A {\bf 56}, 1414 (1997).	
	
\bibitem{CastinDum98}
	Y. Castin and R. Dum,
	Phys. Rev. A {\bf 57}, 3008 (1998).

\bibitem{Leggett01} 
	A. J. Leggett,
	Rev. Mod. Phys. {\bf 73}, 307 (2001).
	
\bibitem{PethickSmith08} 
	C. J. Pethick and H. Smith,
	{\em Bose-Einstein Condensation in Dilute Gases\/} 
	(Cambridge University Press, Cambridge, Second Edition 2008).		
	
\bibitem{PitaevskiiStringari03} 
	L. Pitaevskii and S. Stringari,
	{\em Bose-Einstein Condensation\/}
	(Clarendon Press, Oxford, 2003).
	
\bibitem{Leggett00}
	A. J. Leggett, in
	{\em Connectivity and Superconductivity\/}.
	Lecture Notes in Physics {\bf 62}, 230
	(Springer, Berlin Heidelberg, 2000).

\bibitem{London64}	
	F. London,
	{\em Superfluids. Volume II: Macroscopic Theory of Superfluid Helium\/}
	(Dover, New York, 1964). 
		 
\bibitem{Fock32}
	V. Fock, 
	Z. Phys. {\bf 75}, 622 (1932).
		
\bibitem{GirardeauArnowitt59}
	M. Girardeau and R. Arnowitt,
	Phys. Rev. {\bf 113}, 755 (1959).	
	
\bibitem{Girardeau98}
	M. D. Girardeau,
	Phys. Rev. A {\bf 58}, 775 (1998).

\bibitem{GardinerMorgan07}	
	S. A. Gardiner and S. A. Morgan,
	Phys. Rev. A {\bf 75}, 043621 (2007).	
		
\bibitem{Schrodinger26} 
	E. Schr\"odinger,
	Naturwissenschaften {\bf 14}, 664 (1926).
	
\bibitem{Schiff68}
	L. I. Schiff,
	{\em Quantum Mechanics\/}
	(McGraw-Hill, New York, Third Edition 1968). 
	
\bibitem{KohlerBurnett02}
	Th. K\"ohler and K. Burnett,
	Phys. Rev. A {\bf 65}, 033601 (2002).	 	
	
\bibitem{ErdosEtAl06}
	L. Erd\H{o}s, B. Schlein, and H.-T.\ Yau,
	Comm. Pure Appl. Math. {\bf 59}, 1659 (2006).	
					
\bibitem{ErdosEtAl07} 
	L. Erd\H{o}s, B. Schlein, and H.-T.\ Yau,
	Phys. Rev. Lett. {\bf 98}, 040404 (2007).
	
\bibitem{ErdosEtAl09} 
	L. Erd\H{o}s, B. Schlein, and H.-T.\ Yau,
	J. Amer. Math. Soc. {\bf 22}, 1099 (2009).
		
\bibitem{ErdosEtAl10} 
	L. Erd\H{o}s, B. Schlein, and H.-T.\ Yau,		
	Ann. of Math. {\bf 172}, 291 (2010).
	
\bibitem{LaLifIX}
	E. M. Lifshitz and L. P. Pitaevskii,	
	{\em Statistical Physics, Part~2\/}.
	Vol.~9 of the Landau and Lifshitz Course of Theoretical Physics, \S~26
	(Butterworth-Heinemann, Oxford, 2002).
	
\bibitem{ZhangEtAl04}
	C. Zhang, J. Liu, M. G. Raizen, and Q. Niu,
	Phys. Rev. Lett. {\bf 92}, 054101 (2004).
	
\bibitem{LiuEtAl06}
	J. Liu, C. Zhang, M. G. Raizen, and Q. Niu,
	Phys. Rev. A {\bf 73}, 013601 (2006).
				
\bibitem{DuffyEtAl04}
	G. J. Duffy, A. S. Mellish, K. J. Challis, and A. C. Wilson,
	Phys. Rev. A {\bf 70}, 041602(R) (2004).
	
\bibitem{WimbergerEtAl05}
	S. Wimberger, R. Mannella, O. Morsch, and E. Arimondo,
	Phys. Rev. Lett. {\bf 94}, 130404 (2005).
	
\bibitem{BillamGardiner12}
	T. P. Billam and S. A. Gardiner,
	New J. Phys. {\bf 14}, 013038 (2012).

\bibitem{BillamEtAl13}			
	T. P. Billam, P. Mason, and S. A. Gardiner,
	Phys. Rev. A {\bf 87}, 033628 (2013).		
	
\bibitem{PenroseOnsager56} 
	O. Penrose and L. Onsager,
	Phys. Rev. {\bf 104}, 576 (1956). 
	 	
\bibitem{WeissEtAl05} 
	C. Weiss, S.-A. Biehs, A. Eckardt, and M. Holthaus,
	Laser Physics {\bf 15}, 626 (2005).
	
\bibitem{GertjerenkenHolthaus15}		 	
	B. Gertjerenken and M. Holthaus,
	arXiv:1507.07533.
		
\bibitem{LipkinEtAl65} 
	H. J. Lipkin, N. Meshkov, and A. J. Glick,
	Nuc. Phys. {\bf 62}, 188 (1965).

\bibitem{MeshkovEtAl65} 
	N. Meshkov, A. J. Glick, and H. J. Lipkin,
	Nuc. Phys. {\bf 62}, 199 (1965).

\bibitem{GlickEtAl65} 
	A. J. Glick, H. J. Lipkin, and N. Meshkov,
	Nuc. Phys. {\bf 62}, 211 (1965).
	
\bibitem{GatiMKO07} 
	R. Gati and M. K. Oberthaler,
	J. Phys. B: At. Mol. Opt. Phys. {\bf 40}, R61 (2007).		
				
\bibitem{MilburnEtAl97} 
	G. J. Milburn, J. Corney, E. M. Wright, and D. F. Walls,
	Phys. Rev. A {\bf 55}, 4318 (1997).

\bibitem{ParkinsWalls98} 
	A. S. Parkins and D. F. Walls,
	Phys. Rep. {\bf 303}, 1 (1998).
	
\bibitem{MahmudEtAl05}
	K. W. Mahmud, H. Perry, and W. P. Reinhardt,
	Phys. Rev. A {\bf 71}, 023615 (2005).

\bibitem{BoukobzaEtAl09}
	E. Boukobza, M. Chuchem, D. Cohen, and A. Vardi,
	Phys. Rev. Lett. {\bf 102}, 180403 (2009).

\bibitem{JuliaDiazEtAl10}
	B. Juli\'a-D\'iaz, D. Dagnino, M. Lewenstein, J. Martorell, 
	and A. Polls,
	Phys. Rev. A {\bf 81}, 023615 (2010).
	
\bibitem{NissenKeeling10}
	F. Nissen and J. Keeling,
	Phys. Rev. A {\bf 81}, 063628 (2010).
	
\bibitem{ChuchemEtAl10}
	M. Chuchem, K. Smith-Mannschott, M. Hiller, T. Kottos, A. Vardi,
	and D. Cohen,
	Phys. Rev. A {\bf 82}, 053617 (2010).
	
\bibitem{SimonStrunz12} 
	L. Simon and W. T. Strunz,
	Phys. Rev. A {\bf 86}, 053625 (2012).
	
\bibitem{GraefeEtAl14}
	E.-M. Graefe, H. J. Korsch, and M. P. Strzys,
	J. Phys. A: Math. Theor. {\bf 47}, 085304 (2014).			
			
\bibitem{EilbeckEtAl85} 
	J. C. Eilbeck, P. S. Lomdahl, and A. C. Scott,
	Physica D {\bf 16}, 318 (1985).
	
\bibitem{KenkreCampbell86} 
	V. M. Kenkre and D. K. Campbell,
	Phys. Rev. B {\bf 34}, 4959 (1986).
	
\bibitem{RaghavanEtAl99} 
	S. Raghavan, A. Smerzi, S. Fantoni, and S. R. Shenoi,
	Phys. Rev. A {\bf 59}, 620 (1999).
	
\bibitem{HolthausStenholm01} 
	M. Holthaus and S. Stenholm,
	Eur. Phys. J. B {\bf 20}, 451 (2001).
	
\bibitem{WeissTeichmann08}
	C. Weiss and M. Teichmann, 
	Phys. Rev. Lett. {\bf 100}, 140408 (2008).
	
\bibitem{WeissTeichmann09}
	C. Weiss and M. Teichmann,
	J. Phys. B: At. Mol. Opt. Phys. {\bf 42}, 031001 (2009).						

\bibitem{GertjerenkenHolthaus14} 
	B. Gertjerenken and M. Holthaus,
	New J. Phys. {\bf 16}, 093009 (2014). 
	
\bibitem{ArecchiEtAl72}
	F. T. Arecchi, E. Courtens, R. Gilmore, and H. Thomas,
	Phys. Rev. A {\bf 6}, 2211 (1972).
	
\bibitem{ShampineGordon75}
	L. F. Shampine and M. K. Gordon,
	{\em Computer Solution of Ordinary Differential Equations\/}
	(Freeman and Company, San Francisco, 1975).
	
\bibitem{WrightEtAl96}
	E. M. Wright, D. F. Walls, and J. C. Garrison,
	Phys. Rev. Lett. {\bf 77}, 2158 (1996).
	
\bibitem{LewensteinYou96} M. Lewenstein and L. You,
	Phys. Rev. Lett. {\bf 77}, 3489 (1996).
		
\bibitem{BaronePaterno82}
	A. Barone and G. Paterno,
	{\em Physics and Applications of the Josephson Effect\/}
	(Wiley, New York, 1982).	
		
\bibitem{Gutzwiller90}
	M. C. Gutzwiller,
	{\em Chaos in Classical and Quantum Mechanics\/}
	(Springer-Verlag, New York, 1990).	
		
\bibitem{JoseSaletan98} 
	J. V. Jos\'{e} and E. J. Saletan,
	{\em Classical Dynamics: A Contemporary Approach\/}
	(Cambridge University Press, Cambridge, 1998).
		
\bibitem{CastinDum97}
	Y. Castin and R. Dum,
	Phys. Rev. Lett. {\bf 79}, 3553 (1997).
		
\bibitem{GardinerEtAl00}
	S. A. Gardiner, D. Jaksch, R. Dum, J. I. Cirac, and P. Zoller,
	Phys. Rev. A {\bf 62}, 023612 (2000).
	
\bibitem{BrezinovaEtAl11}
	I. B\v{r}ezinov\'{a}, L. A. Collins, K. Ludwig, B. I. Schneider, 
	and J. Burgd\"orfer,
	Phys. Rev. A {\bf 83}, 043611 (2011).
	
\bibitem{BrezinovaEtAl12}
	I. B\v{r}ezinov\'{a}, A. U. J. Lode, A. I. Streltsov, O. E. Alon,
	L. S. Cederbaum, and J. Burgd\"orfer,
	Phys. Rev. A {\bf 86}, 013630 (2012). 
	
\bibitem{EckardtEtAl05}
	A. Eckardt, C. Weiss, and M. Holthaus,
	Phys. Rev. Lett. {\bf 95}, 260404 (2005).	
	
\bibitem{EckardtEtAl09}
	A. Eckardt, M. Holthaus, H. Lignier, A. Zenesini, D. Ciampini,
	O. Morsch, and E. Arimondo,
	Phys. Rev. A {\bf 79}, 013611 (2009).
	
\bibitem{ZenesiniEtAl09}
	A. Zenesini, H. Lignier, D. Ciampini, O. Morsch, and E. Arimondo,
	Phys. Rev. Lett. {\bf 102}, 100403 (2009).
	
\bibitem{StruckEtAl11}
	J. Struck, C. \"Olschl\"ager, R. Le Targat, P. Soltan-Panahi, 
	A. Eckardt, M. Lewenstein, P. Windpassinger, and K. Sengstock,
	Science {\bf 333}, 996 (2011).
	
\bibitem{MaEtAl11}
	R. Ma, M. E. Tai, P. M. Preiss, W. S. Bakr, J. Simon, and M. Greiner,
	Phys. Rev. Lett. {\bf 107}, 095301 (2011).
	
\bibitem{ChenEtAl11}
	Y.-A. Chen, S. Nascimb\`ene, M. Aidelsburger, M. Atala, S. Trotzky,
	and I. Bloch, 
	Phys. Rev. Lett. {\bf 107}, 210405 (2011).		
	
\bibitem{HaukeEtAl12}
	P. Hauke, O. Tieleman, A. Celi, C. \"Olschl\"ager, J. Simonet,
	J. Struck, M. Weinberg, P. Windpassinger, K. Sengstock, 
	M. Lewenstein, and A. Eckardt,
	Phys. Rev. Lett. {\bf 109}, 145301 (2012).
	
\bibitem{AidelsburgerEtAl13}
	M. Aidelsburger, M. Atala, M. Lohse, J. T. Barreiro, B. Paredes,
	and I. Bloch,
	Phys. Rev. Lett. {\bf 111}, 185301 (2013).				
			
\bibitem{StruckEtAl13}
	J. Struck, M. Weinberg, C. \"Olschl\"ager, P. Windpassinger, 
	J. Simonet, K. Sengstock, R. H\"oppner, P. Hauke, A. Eckardt, 
	M. Lewenstein, and L. Mathey,
	Nature Physics {\bf 9}, 738 (2013).
	
\bibitem{ParkerEtAl13}
	C. V. Parker, L.-C. Ha, and C. Chin,
	Nature Physics {\bf 9}, 769 (2013).	
		
\bibitem{GoldmanDalibard14}
	N. Goldman and J. Dalibard,
	Phys. Rev. X {\bf 4}, 031027 (2014).	
		
\bibitem{Eckardt15}
	A. Eckardt, in preparation.	
			
\end{thebibliography}
\end{document}